\documentclass[a4paper]{amsart}

\usepackage[square,sort,comma,numbers]{natbib}
\usepackage[utf8]{inputenc}
\usepackage[english]{babel}
\usepackage{array}
\usepackage{bm} 
\usepackage{amstext,amssymb,mathtools,esint,tabularx}
\usepackage{natbib}
\usepackage{microtype} 
\usepackage{color, graphicx}
\mathtoolsset{showonlyrefs} 
\usepackage{a4wide}
\usepackage{comment}
\usepackage{fixltx2e} 
 \usepackage{algorithm}
 \usepackage{algpseudocode}

\usepackage{tikz,pgfplots,pgfplotstable}
\usetikzlibrary{calc,3d,arrows,patterns,decorations}
\tikzset{>=stealth'}
\usetikzlibrary{positioning} 
\usepackage[abs]{overpic}

\newtheorem{thm}{Theorem}

\newtheorem{rem}[thm]{Remark}

\def \Vvar {$V_\text{\hspace*{-0.3em}\tiny var}$}

\begin{document}

\title{A quantitative performance analysis for Stokes solvers  at the extreme scale}
\author{Bj\"orn Gmeiner$^\MakeLowercase{a}$ \and Markus Huber$^\MakeLowercase{b}$ \and Lorenz John$^\MakeLowercase{b}$ \and Ulrich R\"ude$^\MakeLowercase{a}$ \and Barbara Wohlmuth$^\MakeLowercase{b}$}
\maketitle
\vspace*{-1.5em}
\begin{center}
\begin{footnotesize}
$^a$ Institute of System Simulation, University Erlangen-Nuremberg, 91058 Erlangen, Germany\\
$^b$ Institute for Numerical Mathematics, Technische Universit\"at M\"unchen, 85748 Garching b.~M\"unchen, Germany
\end{footnotesize}
\end{center}

\begin{abstract}
This article presents a systematic quantitative performance analysis for large finite element computations on extreme scale computing systems. Three parallel iterative solvers for the Stokes system, discretized by low order tetrahedral elements, are compared with respect to their numerical efficiency and their scalability running on up to $786\,432$ parallel threads. A genuine multigrid method for the saddle point system using an Uzawa-type smoother provides the best overall performance with respect to memory consumption and time-to-solution. The largest system solved on a Blue Gene/Q system has more than ten trillion ($1.1 \cdot 10 ^{13}$) unknowns and requires about 13 minutes compute time. Despite the matrix free and highly optimized implementation, the memory requirement for the solution vector and the auxiliary vectors is about 200 TByte. Brandt's notion of ``textbook multigrid efficiency'' is employed to study the algorithmic performance of iterative solvers. A recent extension of this paradigm to ``parallel textbook multigrid efficiency'' makes it possible to assess also the efficiency of parallel iterative solvers for a given hardware architecture in absolute terms. The efficiency of the method is demonstrated for simulating incompressible fluid flow in a pipe filled with spherical obstacles.
\end{abstract}

\medskip
\begin{footnotesize}
\end{footnotesize}


\section{Introduction}\label{Sec:intro}

Current state of the art supercomputers perform several  petaflop per second and 
thus enable large-scale simulations of physical phenomena. 
The enormous computational power is particularly 
valuable for the simulation of 
fluids 
when large computational domains 
must be resolved with a fine mesh.
As an example we will consider the incompressible fluid flow in a pipe filled with spherical obstacles, similar to a 
packed bed, see, e.g., \cite{bogner-mohanty-ruede_2015, dixon_2004, behr_2014}.
A fully resolved mesh  
representing the fluid space between obstacles with small radii 
can lead to discrete 
systems with up to $10^{13}$ degrees of freedom (DoFs).
Grand challenge computations
of such size are only possible with solution techniques 
that are algorithmically optimal and that are specifically
adapted for parallel system architectures.
For minimizing the resource consumption in, e.g., compute time, processor numbers,
memory, or energy, we must proceed beyond the demonstration of asymptotic optimality for 
the algorithms and beyond establishing the scalability of the software.
Modern fast solvers must additionally be implemented with data structures 
that permit an efficient execution on multicore architectures
exploiting their instruction level parallelism. 
They must be adapted to the complex hierarchical memory structure.
This necessitates a meticulous efficiency-aware software design and
an advanced performance aware implementation technology. 
The goal of this article is to provide a systematic performance comparison
of parallel iterative algorithms for the Stokes system
and to explore the current frontiers 
for the size of the systems that can be solved. 

A discretization 
of the underlying partial differential equations with finite elements
has the advantage of being flexible with respect to geometry representation and
different mesh sizes.
Adaptive techniques may become relevant, see,
e.g., \cite{bangerth-burstedde-heister-kronbichler_2011, burstedde-stadler-alisic-wilcox-tan-gurnis-ghattas_2013, burstedde-wilcox-ghattas_2011, stals_1996}.
Nevertheless, these methods require dynamic data structures that can lead to 
a significant computational cost and parallel overhead if they are not handled properly.
Moreover, high-order methods are found to be attractive 
since they may reach a higher 
accuracy with the same number of degrees of freedom, 
but they will in turn lead to denser matrices. 
Consequently, they require more memory, which will also translate into 
more memory traffic and is thus often the limiting factor on current
computer architectures \cite{hager2010introduction}. 
Recent contributions on high-order methods for massively parallel architectures can be found,
e.g., in \cite{sundar-stadler-biros_2015}.
Realizing scalable finite element methods is only possible on the basis of asymptotically optimal solvers, such as multigrid methods, whose complexity grows linearly with the problem size.

In this article, we study the performance of different, existing solution algorithms and their influence on different formulations of the Stokes system, which include the Laplace- or the divergence of the symmetric part of the gradient operator. Iterative solution techniques for saddle point problems have a long history, see, e.g., \cite{bank-welfert-yserentant_1990,  benzi-golub-liesen_2005, brandt1982guide, brandt-dinar_1979, elman-silvester-wathen_2005, maitre-musy-nigon_1985, verfuerth_1984}. In general, we distinguish between preconditioned iterative methods (Krylov subspace methods) and genuine all-at-once multigrid methods where the multigrid algorithm is used as a solver for the saddle point problem. Both solver types are studied within this article and numerical examples illustrate that they perform with excellent scalability on massively parallel machines. In case of the all-at-once multigrid method we reach up to $10^{13}$ DoFs.

These solution techniques are realized 
in the hierarchical hybrid grids framework (HHG) \cite{bergen-huelsemann_2003, bergen-huelsemann_2004}. HHG presents a compromise between
structured and unstructured grids.
It exploits the flexibility of finite elements and capitalizes on the parallel efficiency of geometric multigrid methods.
The HHG package has been designed originally
as multigrid library for elliptic problems, and its excellent
efficiency was demonstrated for scalar equations in
\cite{bergen-gradl-ruede-huelsemann_2006, bergen20051}. 
In \cite{gmeiner-ruede-stengel-waluga-wohlmuth_2015, gmeiner-ruede-stengel-waluga-wohlmuth_2015_2}
an extension of HHG to solve the Stokes system can be found.
Here, we will extend this work further and will lay a 
particular emphasis on the robustness and efficiency (time-to-solution). 

This article is structured as follows: In Sec.~\ref{sec:modeldiscr}, we introduce the model problem and a low order, stabilized finite element discretization for the Stokes system. The HHG framework and the solution algorithms will be introduced in Sec.~\ref{sec:solver}. These include (i) an approximate Schur-complement CG algorithm for the pressure, (ii) a multigrid preconditioned MINRES method,  and (iii) an all-at-once multigrid method that uses Uzawa-type smoothers.  A detailed comparison of the solvers will be presented in Sec.~\ref{sec:comparison}, which also illustrates the performance of the methods for the serial and the massively parallel case. In particular, we demonstrate that already on a conventional low cost workstation we can solve systems with $5 \cdot 10^8$ DoFs and on large petascale systems up to $10^{13}$ DoFs, both, with a compute time below 13 minutes. Moreover, we present in Sec.~\ref{sec:tme} an extension of the textbook multigrid efficiency concept, introduced in \cite{brandt1998barriers,gmeiner-ruede-stengel-waluga-wohlmuth_2015_2}, for the Uzawa-type multigrid method. Finally, in Sec.~\ref{sec:appli}, we show an application to the incompressible fluid flow in a pipe filled with spherical obstacles with different types of boundary conditions.

\section{Model problem and discretization}\label{sec:modeldiscr}

For the simulation of incompressible fluid flow at low Reynolds number, it is sufficient to describe the physics by the Stokes system, where additional nonlinear terms are neglected. Thus, we consider the Stokes system in the following as a model problem, in a polyhedral domain $\Omega \subset \mathbb{R}^3$ and for simplicity with homogenous Dirichlet boundary conditions on the boundary $\Gamma = \partial \Omega$, see Sec.~\ref{sec:appli} for the more general case of Neumann  and free-slip boundary conditions. The balance equations for mass and momentum then read as
\begin{equation}\label{eq:stokes}
\begin{alignedat}{3}
-\mathop{\rm div} T(\mathbf{u}, p) &= \mathbf{f} &\qquad& \text{in } &&\Omega,\\
\mathop{\rm\,div}\mathbf{u} &= 0 && \text{in } &&\Omega, \\
\mathbf{u} &= \mathbf{0} && \text{on } &&\Gamma.
\end{alignedat}
\end{equation}
Here, $\mathbf{u}$ denotes the fluid velocity, $p$ the pressure, and $\mathbf{f} \in L^2(\Omega)^3$ a given forcing term, acting on the fluid. Further, we denote by 
\begin{align}\label{cauchystress}
T(\mathbf{u}, p) \coloneqq 2 \nu D(\mathbf{u}) - p I,
\end{align}
the Cauchy stress tensor, which includes the viscosity constant $\nu > 0$, and the symmetric part of the velocity gradient
\begin{align*}
D(\mathbf{u}) \coloneqq \frac{1}{2}(\nabla \mathbf{u} + (\nabla \mathbf{u})^\top).
\end{align*}
We refer for instance to \cite{pironneau_1989} for a more detailed study on the modeling part.
\begin{rem}
Note that due to the constant viscosity $\nu \in \mathbb{R}_+$, the momentum equation of the Stokes system \eqref{eq:stokes} can be equivalently written as 
\begin{align*}
-\nu \Delta \mathbf{u}+ \nabla{p} = \mathbf{f} \qquad \textnormal{in } \Omega.
\end{align*}
\end{rem}
In order to reflect the physics correctly, the symmetric part of the velocity gradient $D(\mathbf{u})$ is particularly needed in the case of varying viscosities or when the natural boundary condition $T(\mathbf{u}, p)\, \mathbf{n} = \mathbf{0}$ is considered. Here, we consider the constant viscosity case, and thus two formulations with the Laplace- and $D$-operator are possible. 

In both cases, the pressure is only well-defined up to an additive constant, thus we exclude the constants and introduce the space
\begin{align*}
L_0^2(\Omega) = \{ q \in L^2(\Omega)\, :\, \langle q, 1 \rangle_\Omega = 0 \},
\end{align*}
for the pressure, where $\langle \cdot, \cdot \rangle_\Omega$ denotes the inner product in $L^2(\Omega)$. The existence and uniqueness of a solution of the Stokes system \eqref{eq:stokes} is well understood, see, e.g., \cite{brezzi-fortin_book, elman-silvester-wathen_2005, girault-raviart_book}.

\subsection{Finite element discretization}
The computational domain is subdivided into a conforming tetrahedral initial mesh $\mathcal{T}_{-2}$ with possibly unstructured and anisotropic elements. Based on this initial triangulation, we construct a hierarchy of grids $\mathcal{T}\coloneqq\{\mathcal{T}_\ell,\, \ell=0,1,\dots,L\}$ by successive uniform refinement, conforming to the array-based data structures used in the hierarchical hybrid grids (HHG) implementation, cf. Sec.~\ref{sec:hhg_framework} and \cite{bergen-huelsemann_2004, gmeiner_diss} for details. We point out that this uniform refinement strategy guarantees that all our meshes satisfy a uniform shape-regularity.

For the discretization, we use linear, conforming finite elements, i.e., for a given mesh $\mathcal{T}_\ell \in \mathcal{T}$, $\ell \in \mathbb{N}_0$, we define the function space of piecewise linear and globally continuous functions by
\begin{align*}
S^1_\ell (\Omega) \coloneqq \{ v \in \mathcal{C}(\overline{\Omega}) :  v|_T \in P_1(T),~\forall \, T \in \mathcal{T}_\ell \}.
\end{align*}
The conforming, finite element spaces for velocity and pressure are then given by
\begin{align}\label{fespaces}
\mathbf{V}_\ell = [S^1_\ell(\Omega) \cap H^1_0(\Omega)]^3, \qquad Q_\ell = S^1_\ell(\Omega) \cap L^2_0(\Omega).
\end{align}
The boundary conditions for the velocity $\mathbf{u}$ as well as the mean-value condition of the pressure $p$ are directly built into the finite element spaces. Since the above finite element pair \eqref{fespaces} does not satisfy the discrete inf--sup condition, we need to stabilize the method. Within this work we consider the standard PSPG stabilization, see, e.g., \cite{brezzi-douglas_1988}. The extension to other stabilizations, such as the Dohrmann--Bochev approach \cite{bochev-dohrmann_2004} or the local projection stabilization \cite{ganesan-matthies-tobiska_2008}, can be done similarly. This leads to the following two discrete variational formulations of the Stokes system \eqref{eq:stokes}. Find $(\mathbf{u}_\ell,p_\ell) \in \mathbf{V}_\ell \times Q_\ell$ such that
\begin{equation}\label{eq:varform}
\begin{alignedat}{4}
&a_i(\mathbf{u}_\ell,\mathbf{v}_\ell) &&+ b(\mathbf{v}_\ell,p_\ell) &&= f(\mathbf{v}_\ell)  &\qquad&\forall\,\mathbf{v}_\ell \in \mathbf{V}_\ell,\\
&b(\mathbf{u}_\ell,q_\ell) &&- c_\ell(q_\ell,p_\ell) &&= g_\ell(q_\ell) && \forall\,q_\ell \in Q_\ell,
\end{alignedat}
\end{equation}
for $i=1,2$, where the bilinear and linear forms are given by
\begin{alignat}{2}
a_1(\mathbf{u},\mathbf{v}) &\coloneqq \langle \nu \nabla \mathbf{u}, \nabla \mathbf{v} \rangle_{\Omega}, \quad
a_2(\mathbf{u},\mathbf{v}) &&\coloneqq 2\langle \nu D(\mathbf{u}), D(\mathbf{v}) \rangle_{\Omega},\\
b(\mathbf{u},q) &\coloneqq -\langle \mathbf{\rm div}\, \mathbf{u}, q \rangle_{\Omega}, \qquad \ \,
f(\mathbf{v}) &&\coloneqq \langle \mathbf{f}, \mathbf{v} \rangle_{\Omega},
\end{alignat}
for all $\mathbf{u}, \mathbf{v} \in H_0^1(\Omega)^3$, $q \in L_0^2(\Omega)$. Furthermore, the level-dependent stabilization terms $c_\ell(\cdot,\cdot)$ and $g_\ell(\cdot)$ are given by
\begin{align*}
c_\ell(q_\ell,p_\ell) \coloneqq \sum_{T \in \mathcal{T}_\ell} \delta_T\, h_T^2 \,\langle \nabla p_\ell, \nabla q_\ell \rangle_T
\quad\text{and}\quad
g_\ell(q_\ell) \coloneqq -\sum_{T \in \mathcal{T}_\ell} \delta_T\, h_T^2 \, \langle \mathbf{f}, \nabla q_\ell \rangle_T,
\end{align*}
with $h_T = (\int_T  dx)^{1/3}$.
The stabilization parameter $\delta_T > 0$ has to be chosen carefully to guarantee uniformly stable equal order approximations and to avoid unwanted effects due to over-stabilization. We fix $\delta_T = 1/12$ which is a good choice in practice, see also \cite{elman-silvester-wathen_2005}.

In what follows, we shall refer to the formulation \eqref{eq:varform} with $a_1(\cdot,\cdot)$ as Laplace-operator formulation and with $a_2(\cdot,\cdot)$ as $D$-operator formulation.

\begin{rem}
The discrete Laplace- and $D$-operator formulations \eqref{eq:varform} are not equivalent, both lead to different numerical solutions. The discretization errors for velocity and pressure are for both formulations of the same quality and order. Nevertheless, the efficiency of iterative solvers might differ, which will be also reflected in the computing times. Therefore, we shall study both formulations and their comparison.
\end{rem}


\section{Solvers for the Stokes system}\label{sec:solver}

In this section, we briefly discuss the concept of the  HHG framework and recall, for convince of the reader, the characteristic properties of the considered Krylov subspace and all-at-once multigrid methods, see \cite{maitre-musy-nigon_1985, verfuerth_1984, wathen-silvester_1993}.

In the following, we denote by $n_{\ell,u} = {\rm dim}\, \mathbf{V}_\ell$ and $n_{\ell,p} = {\rm dim}\, Q_\ell$ for $\ell = 0, \dots, L$ the number of degrees of freedom for velocity and pressure, respectively. Then, the following isomorphisms $\mathbf{u}_\ell \leftrightarrow \mathbf{u} \in \mathbb{R}^{n_{\ell,u}}$ and $p_\ell \leftrightarrow \mathbf{p} \in \mathbb{R}^{n_{\ell,p}}$ are satisfied, and the algebraic form of the variational formulation \eqref{eq:varform} reads as
\begin{align}\label{eq:linear-system}
\mathcal{K} 
\begin{pmatrix}
\mathbf{u} \\
\mathbf{p}
\end{pmatrix} 
\coloneqq
\begin{pmatrix}
A_i & B^\top \\
B & -C
\end{pmatrix}
\begin{pmatrix}
\mathbf{u} \\
\mathbf{p}
\end{pmatrix} 
=
\begin{pmatrix}
\mathbf{f} \\
\mathbf{g}
\end{pmatrix}
\eqqcolon \tilde{\mathbf{f}},
\end{align}
with the system matrix $\mathcal{K} \in \mathbb{R}^{(n_{\ell,u} + n_{\ell,p}) \times (n_{\ell,u} + n_{\ell,p})}$. The matrix $A_i \in \mathbb{R}^{n_{\ell,u} \times n_{\ell,u}}$, corresponding to $a_i(\cdot, \cdot)$, $i=1,2$, itself consist of a $3\times3$ block structure, where the blocks are in general non-zero, particularly for the $D$-operator formulation. However, in the case of the Laplace-operator formulation we have only 3 nontrivial blocks, given by the stiffness matrices of the Laplacian, on the diagonal. This has two advantages. First, the system matrix $\mathcal{K}$ is more sparse and second, the solver does not need to take into account the cross coupling terms.

\subsection{Schur-complement CG -- pressure correction scheme}

Starting from a code for scalar elliptic equations, the most natural approach to deal with the Stokes system is the Schur-complement conjugate-gradient (CG) algorithm, see, e.g., \cite{verfuerth_1984}, also known as pressure correction scheme. This method offers a possibility to reuse algorithms for symmetric and positive definite algebraic systems.
By formally eliminating the velocities from the pressure equation, we can reformulate
 \eqref{eq:linear-system} equivalently as
\begin{align*}
\label{eq:stokes-discrete-system}
\begin{pmatrix}
A_i & B^\top\\
0 & BA_i^{-1}B^\top + C\\
\end{pmatrix}
\begin{pmatrix}
\mathbf{u}\\
\mathbf{p}
\end{pmatrix}
&=
\begin{pmatrix}
\mathbf{f}\\
BA_i^{-1}\mathbf{f} - \mathbf{g}
\end{pmatrix}.
\end{align*}
For the pressure, we obtain the Schur-complement equation as 
\begin{equation}\label{eq:pressure-schur}
S_i\mathbf{p} = \mathbf{r},
\end{equation}
where $S_i \coloneqq BA_i^{-1}B^{\top} + C$ denotes the Schur-complement and $\mathbf{r} =BA_i^{-1}\mathbf{f} - \mathbf{g}$. The Schur-complement equation \eqref{eq:pressure-schur} is solved by a preconditioned CG method, where we choose a lumped mass matrix $M$ as preconditioner, since it is spectrally equivalent, i.e.
\begin{align*}
(S_i \mathbf{q}, \mathbf{q}) \simeq (M \mathbf{q}, \mathbf{q}),
\end{align*}
for all $\mathbf{q} \in \mathbb{R}^{n_{\ell, p}}$, $\mathbf{q} \leftrightarrow q_\ell \in Q_\ell$, see, e.g., \cite{elman-silvester-wathen_2005, grinevich-olshanskii_2009}.

Since the direct assembly of the Schur-complement matrix $S_i$ cannot be performed efficiently, it is applied indirectly by replacing each multiplication of the discrete inverse Laplacian $A_i^{-1}$ by a multigrid algorithm. This results, for given iteration numbers $n_A, n_S, n_I \in \mathbb{N}$, in the following Schur-complement CG algorithm:
\bigskip

\noindent \textbf{Algorithm 1:}\\
For $k = 1, \dots$

\begin{itemize}
\item Solve the equation
\begin{align}
A_i \mathbf{u}_{k} = \mathbf{f} - B^\top \mathbf{p}_{k-1}
\end{align}
by applying $n_A$ iterations of a multigrid method.
\smallskip
\item Do $n_S$ CG iterations for the Schur-complement equation, preconditioned by a lumped mass matrix, to approximate 
\begin{align}
S_i\mathbf{p}_{k} = \mathbf{r},
\end{align}
where the start iterate $\mathbf{p}_{k-1}$ and the initial residual $\mathbf{r}_{k-1} = B \mathbf{u}_{k} - C \mathbf{p}_{k-1} + \mathbf{g}$ are used. Further, within the CG algorithm, $A_i^{-1}$ is approximated by $n_I$ iterations of a multigrid method.
\end{itemize}

\subsection{The preconditioned MINRES method}\label{sec:pminres}

Since the linear system \eqref{eq:linear-system} is symmetric and indefinite, a conjugate gradient method cannot be used as an iterative solver for the overall system. Nevertheless, the minimal residual method (MINRES) is applicable which requires in general a symmetric preconditioner, see, e.g., \cite{elman-silvester-wathen_2005, wathen-silvester_1993}. The system matrix $\mathcal{K}$ in \eqref{eq:linear-system} is spectrally equivalent to the block diagonal preconditioner
\begin{equation}\label{precond_block_diag}
\begin{aligned}
\begin{pmatrix}
\hat{A}_i & 0 \\
0 & \hat{S}_i
\end{pmatrix},
\end{aligned}
\end{equation}
where $\hat{A}_i$ and $\hat{S}_i$ denote spectral equivalent preconditioners for $A_i$ and the Schur-complement $S_i$, respectively, see, e.g., \cite{elman-silvester-wathen_2005, wathen-silvester_1993}. Again, $\hat{A}_i^{-1}$ will be efficiently realized by a multigrid method and $\hat{S}_i^{-1}$ by a lumped mass matrix $M$. 

Alternatively, one may think of different symmetric preconditioners, motivated from a matrix factorization, see, e.g., \cite{bank-welfert-yserentant_1990, klawonn_1998}.
Note, that such preconditioners result in an algorithm where we need to solve two times for the velocity and thus one iteration is more expensive than of the simple block diagonal preconditioner \eqref{precond_block_diag}. On the other hand we may have fewer overall iterations, but from our experience the simple block diagonal preconditioner \eqref{precond_block_diag} is more efficient with respect to time-to-solution and thus chosen in the forthcoming numerical examples.

\subsection{Uzawa-type multigrid methods}

Using multigrid directly to solve the Stokes system, has already been done in the early multigrid development, see, e.g., \cite{brandt1982guide, maitre-musy-nigon_1985}. As a third solver we therefore consider a multigrid method for the indefinite system \eqref{eq:linear-system}. The key point for such a method is the construction of a suitable smoother for the saddle point problem. Several different classes of smoothers are available, such as the Vanka smoothers \cite{vanka_1986, wobker-turek_2009}, the Braess--Sarazin smoother \cite{braess-sarazin_1997}, distributive smoothers \cite{brandt-dinar_1979}, or transforming smoothers \cite{wittum_1989, wittum_1990}, see also \cite{wieners_2000}. Also, Uzawa-type smoothers have been successfully constructed and applied, see, e.g., \cite{bank-welfert-yserentant_1990, gaspar2014simple, schoeberl-zulehner_2003, zulehner_2002}. Note, the Uzawa-type algorithm is here applied locally as a smoother, while the multigrid algorithm frames the solver part. 
Uzawa-type multigrid methods require only nearest neighbor communication 
like scalar point smoothers. 
Different from e.g.\ the distributed smoothers proposed in \cite{brandt1982guide, brandt-dinar_1979},
that require a more complex parallel communication pattern,
Uzawa-type smoothers are easily usable within the HHG framework.

The idea of the Uzawa smoother is based on a preconditioned Richardson iteration. In the $(k+1)$th-iteration, we solve the system
\begin{align*}
\begin{pmatrix}
\hat{A}_i & 0 \\
B & -\hat{S}_i
\end{pmatrix}
\begin{pmatrix}
\mathbf{u}_{k+1} - \mathbf{u}_k\\
\mathbf{p}_{k+1} - \mathbf{p}_k
\end{pmatrix}
=
\begin{pmatrix}
\mathbf{f} \\
\mathbf{g}
\end{pmatrix}
- \mathcal{K}
\begin{pmatrix}
\mathbf{u}_k\\
\mathbf{p}_k
\end{pmatrix},
\end{align*}
for $(\mathbf{u}_{k+1}, \mathbf{p}_{k+1})^\top$, 
where $\hat{A}_i$ and $\hat{S}_i$ denote preconditioners for $A_i$ and the Schur-complement $S_i$, respectively. These will be specified by the smoothers in detail later. The application of the preconditioner results in the algorithm of the \textit{inexact Uzawa method}, where we smooth the velocity part in a first step and in a second step the pressure, i.e.
\begin{equation}\label{Uzawa_smooth}
\begin{aligned}
\mathbf{u}_{k+1} &= \mathbf{u}_k + \hat{A}_i^{-1} (\mathbf{f} - A_i \mathbf{u}_k - B^\top \mathbf{p}_k),\\
\mathbf{p}_{k+1} &= \mathbf{p}_k + \hat{S}_i^{-1} (B \mathbf{u}_{k+1} - C \mathbf{p}_k - \mathbf{g}).
\end{aligned}
\end{equation}
As in Sec.~\ref{sec:pminres}, we may think of a symmetric preconditioner, see, e.g., \cite{schoeberl-zulehner_2003, zulehner_2002}. However, numerical tests show that this alternative variant is not preferable with respect to the time-to-solution and thus will not be further considered in this work. For the convergence analysis of these methods see also \cite{gaspar2014simple, schoeberl-zulehner_2003, zulehner_2002}.

\subsection{The HHG framework and smoothers}\label{sec:hhg_framework}
\begin{figure}[h!]
\centering
\includegraphics[width=0.6\textwidth]{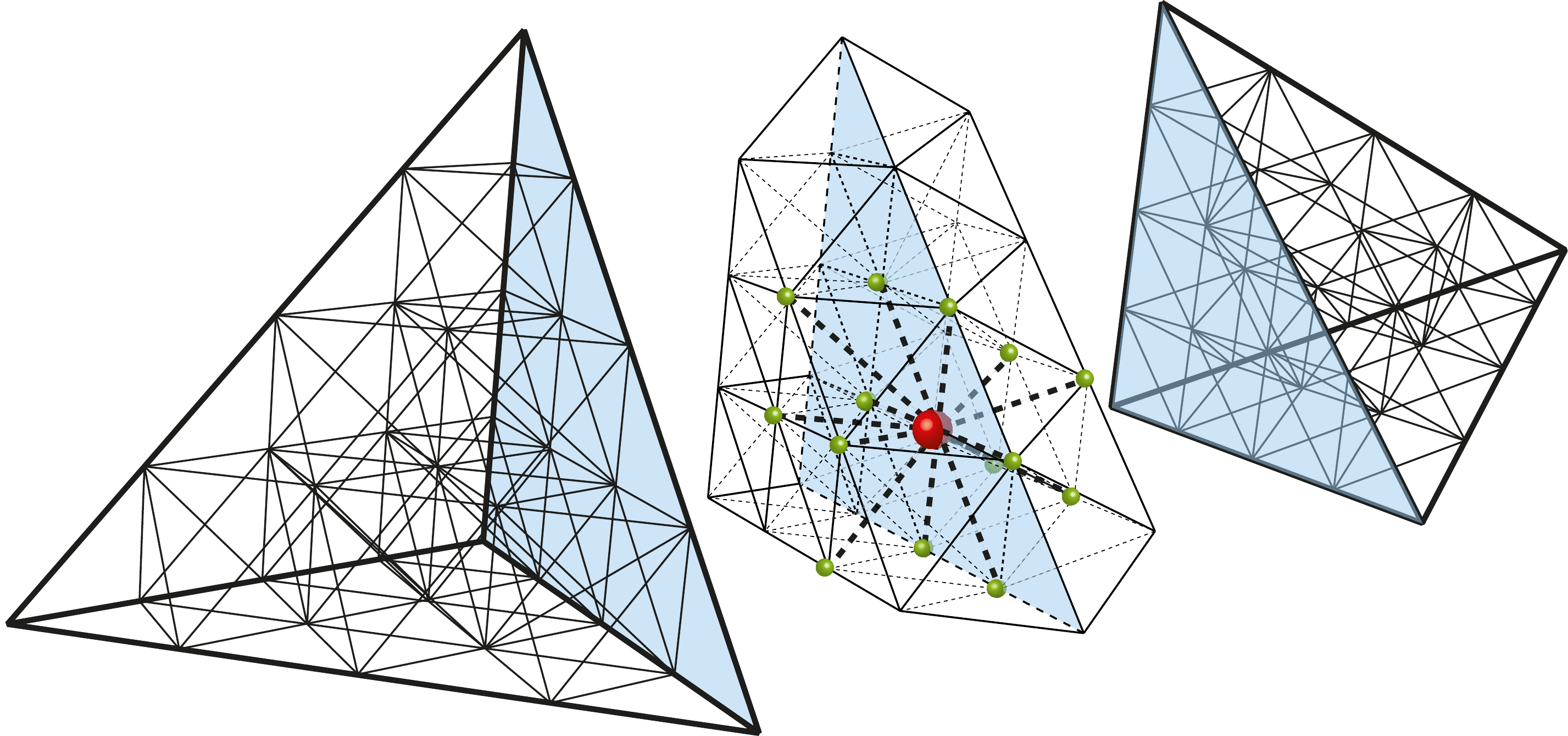}
\caption{Ghost layer structure and 15-point stencil (red/green).}
\label{F:halostructure}
 \end{figure}

The HHG framework is a carefully designed and implemented high performance finite element geometric multigrid software package \cite{bergen-gradl-ruede-huelsemann_2006, bergen-huelsemann_2004}.
HHG combines the flexibility of unstructured finite element
meshes with the performance advantage of structured grids in a block-structured approach.
It employs an unstructured input grid that
is structurally refined. 
The grid is then 
organized into primitive types according to the original input grid structure:
vertices, edges, faces, and volumes.
These primitives become container data structures for the nodal values of the refined mesh. Moreover, an algorithmic traversal of a fine mesh is naturally organized into an iteration over the primitives, and the mesh nodes contained in each of them.
We exploit this data structure in all multigrid operations such as smoothing, prolongation, restriction, and residual calculation. 
Each of these subroutines operates locally on the primitive
itself.
Read access to the nodes that are contained in other primitives 
is accomplished via ghost layers, see Fig.~\ref{F:halostructure} for an
illustration. 

The dependencies between primitives are updated by ghost layer exchanges
in 
an ordering from the vertex primitives via the edges and faces to the volume primitives.
This scheme is designed to support  an efficient parallel computation on 
distributed memory systems and is implemented using message passing with MPI.
Communication is applied only in one way, i.e., copying of data is always
handled by the primitive of higher geometrical order. This design decision imposes a natural ordering of a block Gauss--Seidel 
structure based on the primitive classes. To facilitate the parallelization and reduce the communication, the primitives within each class
are decoupled resulting in a block Jacobi structure. Finally, we are free to specify the smoother acting on the nodes of each
primitive, see, e.g.,  \cite{bergen-gradl-ruede-huelsemann_2006, bergen-huelsemann_2003, bergen-huelsemann_2004}. 
Using locally a Gauss--Seidel smoother for each primitive, we call the associated global smoother a hybrid Gauss--Seidel method. We point out that only ${\mathcal O} (h^{-1})$ nodes  are handled partly in a Jacobi type way, while ${\mathcal O} (h^{-3})$ nodes are updated by a proper Gauss--Seidel scheme. A row-wise red-black coloring within each primitive defines the so-called forward hybrid Gauss--Seidel (FHGS) while the backward variant (BHGS) only reverses the ordering within each primitive but not the ordering in the primitive hierarchy. Thus  the sequential execution  of a FHGS and BHGS smoother does not yield a symmetric operator. From now on, we will refer to this variant as pseudo-symmetric hybrid block Gauss--Seidel smoother (SHGS). A $V$-cycle  with $n_\text{pre}$ pre-smoothing and $n_\text{post}$ post-smoothing steps is denoted by $V(n_\text{pre}, n_\text{post})$.

Note that due to the low order discretization,
only nearest-neighbor communication
is necessary and that this
accelerates the parallel communication. 
Moreover, HHG provides a storage-efficient matrix-free implementation 
that avoids the assembly, storage, and access to the global matrices.
HHG containers are realized as array-based data structures such that
performance critical indirect memory access is systematically avoided. 
With the block-structure, HHG can furthermore exploit a stencil-like
approach which boosts the  performance of computation-intensive matrix-vector operations.
Additionally, optimized compute kernels are available \cite{gmeiner-ruede-stengel-waluga-wohlmuth_2015}.
\subsection{Smoothers and parameters for the solvers}\label{Sec:smoother_param}
Unless specified otherwise, the following smoothers and parameters are applied for the individual solvers.
For the Schur-complement CG algorithm a multigrid iteration 
is required only for the velocity components. In our case, we use a $V(3,3)$--cycle and apply the FHGS scheme within the pre-smoothing steps and the  BHGS for the post-smoothing steps, denoted by (FHGS, BHGS). The number of inner- and outer-iterations, according to  Alg.~1, are specified by $n_A = n_S = 3$ and $n_I = 1$.
Similarly, for the preconditioned MINRES method with block-diagonal preconditioning only a multigrid iteration for the velocity part is needed. Here, we use a $V(1,1)$--cycle with the (FHGS, BHGS) smoother as a preconditioner for the $A_i$-block.

The analysis of the Uzawa-type multigrid method actually requires a $W$--cycle, see, e.g., \cite{schoeberl-zulehner_2003} which is too costly, particularly for the parallel case, due to 
the extensive work on the coarse grid. As an alternative we consider a variable $V$--cycle, which is closer to the $W$--cycle than a standard $V$--cycle but with the advantage of 
a lower coarse grid cost. 
In our case, we employ for the variable $V$--cycle two additional pre- and post-smoothing steps on each level in comparison to the previous level. Still, we need to specify the smoothers for velocity and pressure, cf. \eqref{Uzawa_smooth}. We consider for the velocity part a pseudo-symmetric hybrid Gauss--Seidel (SHGS) smoother. For the pressure, we consider a forward hybrid Gauss--Seidel smoother (FHGS), applied to the stabilization matrix $C$, with under-relaxation $\omega = 0.3$. These smoothers are then applied within a variable $V(3,3)$--cycle, denoted by \Vvar$(3,3)$.

\subsection{Stopping criteria}
Let us denote by $\mathbf{x}_0 = (\mathbf{u}_0, \mathbf{p}_0)^\top$ the initial guess for velocity and pressure, $\mathbf{u}_0$ is generated as a random vector with values in $[0,1]$ and the same for the pressure initial $\mathbf{p}_0$, which is additionally scaled by $h_\text{min}^{-1}$, $h_\text{min} = \min_{T \in \mathcal{T}_\ell} h_T$, i.e. with values in $[0, h_\text{min}^{-1}]$. This is important in order to reflect a less regular pressure, since the velocity is generally considered as a $H^1(\Omega)$ 
function and the pressure as a $L^2(\Omega)$ function. Important, for a fair comparison is the consideration of the same stopping criteria. In fact, choosing the standard stopping criteria of the individual solvers can lead to quite different relative-, absolute residuals and consequently solutions, depending on the specified accuracy. We choose as stopping criterion for all methods the relative residual with a tolerance $\varepsilon = 10^{-8}$, i.e.,
\begin{align*}
\frac{\| \mathcal{K} \mathbf{x}_k - \tilde{\mathbf{f}} \|_2}{\| \mathcal{K} \mathbf{x}_0 - \tilde{\mathbf{f}} \|_2} \leq \varepsilon,
\end{align*}
where $||\cdot ||_2$ denotes the standard Euclidean norm. 
In passing we note that multigrid methods often provide more accurate solutions
for the same residual tolerance, as compared to other iterative methods.
This is caused by the feature of multigrid methods that the residuals are reduced on all
levels of a multi-scale hierarchy, so that the asymptotically
deteriorating conditioning of the system matrix $\mathcal{K}$
has less effect on the accuracy of the solution for a given residual tolerance. 

\subsection{Coarse grid solver}\label{Sec:coarse_grid}
For the solver on the coarsest level, several possibilities are available, this includes parallel sparse direct solvers, such as PARDISO \cite{pardiso2} or algebraic multigrid methods, such as hypre \cite{falgout-jones-yang_2006}. For our current performance study, we want to avoid the dependency on external libraries, and thus simply employ suitable Krylov subspace coarse-grid solvers. 

In the case of the Schur-complement CG algorithm and the preconditioned MINRES method, we only need a coarse grid solver for the $A_i$-block, which motivates to consider a standard CG algorithm. The number of CG iterations will be specified.
However, in the case of the Uzawa-type multigrid method, we have to consider a coarse grid solver for the full algebraic system \eqref{eq:linear-system}. Due to this reason we choose a MINRES method as a coarse grid solver, which is preconditioned in a block diagonal fashion \eqref{precond_block_diag}, with a CG method and lumped mass matrix. Relative accuracies for the Krylov subspace algorithms are specified, i.e., for the preconditioned MINRES method we consider $\varepsilon = 5 \cdot 10^{-3}$ and for the CG method $\varepsilon = 10^{-3}$, which seems to be a good choice.

We summarize the algorithmic set-up for the Schur-complement CG (SCG) algorithm, the preconditioned MINRES (PMINRES) and the Uzawa-type multigrid (UMG) method in Tab.~\ref{tab:settings}.
\begin{table}[h]
\centering
\begin{tabular}{ c | c c c}
\hline
 				& SCG & PMINRES & UMG\\ \hline
$\hat{A}$ 			& (FHGS, BHGS) & (FHGS, BHGS) & SHGS\\ 
$\hat{S}$ 			& $M$ & $M$ & FHGS ($\omega = 0.3$)\\ 
cycle, smooth. steps	& $V(3,3)$ & $V(1,1)$ & \Vvar$(3,3)$ \\ 
coarse grid solver 	& CG & CG & PMINRES\\ 
add. inform. 		& $n_A = n_S = 3$, $n_I = 1$ & -- & --\\ 
\hline
\end{tabular}
\caption{Algorithmic set-up for the individual methods.}
\label{tab:settings}
\end{table}

\section{Comparison of the solvers}\label{sec:comparison}
In this section, we compare the three solvers with respect to time-to-solution, operator evaluations and scalability. Computations are performed for serial and massively parallel test cases with up to $8 \cdot 10^5$ threads. As a computational domain, we consider the unit cube $\Omega = (0,1)^3$.  The initial mesh $\mathcal{T}_{-2}$ for the serial tests consists of 6 tetrahedrons and for the weak-scaling it is adapted to the number of threads. A uniform refinement step decomposes each tetrahedron into 8 new ones. Note, the coarse grid $\mathcal{T}_0$ is a two times refined initial grid. We choose a homogenous right-hand side $\mathbf{f} = \mathbf{0}$ and the viscosity $\nu = 1$. 

Computations will be shown first on a conventional low cost workstation for the serial case with a single Intel Xeon CPU E2-1226 v3, 3.30GHz and 32 GB shared memory. The parallel computations are performed on JUQUEEN (J\"ulich Supercomputing Center, Germany), currently listed as
number 9 of the TOP500 list\footnote{\tt{http://top500.org}, June 2015}.

\subsection{Time-to-solution}\label{timetosol}
For this first numerical example, we consider the Laplace-operator formulation and present iteration numbers (iter) and time-to-solution (time) for the individual solvers. These are performed on a single processor, as specified above. Since the problem size on the fixed coarse grid is relatively small (81 or 206 DoFs, depending on the method), we set the number of coarse grid CG/MINRES iterations to 5.

\begin{table}[h!]
\centering
\begin{tabular}{| l  r | r  r | r  r | r  r |}
\hline
\multicolumn{2}{|c|}{}  & \multicolumn{2}{c|}{SCG} & \multicolumn{2}{c|}{PMINRES} & \multicolumn{2}{c|}{UMG}\\
\hline
$L$ & DoFs & iter & \multicolumn{1}{c|}{time} & iter & \multicolumn{1}{c|}{time} & iter & \multicolumn{1}{c|}{time} \\
\hline
2 &  $1.5 \cdot 10^{4\ }$	& 26	&      0.11	& 108	&     0.23 		&   9 &   0.08 \\ 
3 &  $1.3 \cdot 10^{5\ }$	& 28	&     0.58 	& 83 		&     0.78		&   8 &   0.29 \\ 
4 &  $1.0 \cdot 10^{6\ }$	& 28	&     3.50 	& 73		&     4.05		&   8 &   1.79 \\ 
5 &  $8.2 \cdot 10^{6\ }$	& 28	&   25.48 	& 70		&   26.91		&   8 & 12.70 \\
6 &  $6.6 \cdot 10^{7\ }$	& 31	& 215.03 	& 67		& 192.09 		&   8 & 95.85 \\
7 & $5.3 \cdot 10^{8\ }$	& 29 & 1533.52 & \multicolumn{2}{c|}{out of memory} &  8 & 730.77 \\
\hline
\end{tabular}
\caption{Iteration numbers and time-to-solution (in sec.) for the Laplace-operator.}
\label{T:itertime1}
\end{table}
In Tab.~\ref{T:itertime1}, we present results for several refinement levels $L$. We observe that all three solvers are robust with respect to the mesh size. While the SCG algorithm and the PMINRES method are similar, the UMG method is approximately by a factor of two faster. 

Note that the PMINRES method is running out of memory on refinement level $L = 7$, whereas the SCG algorithm and the UMG method can still solve this system. The UMG method requires 3 vectors, unknowns (velocity and pressure), right-hand side and residual vector on the finest level and no additional auxiliary vectors like in the case of the SCG algorithm and the PMINRES method. 
We recall that the HHG framework uses a matrix-free approach so that only a negligible amount of memory is spent on assembling and storing the matrices.
In order to compute the memory allocation costs of the UMG method, we further neglect ghost layer and coarse grid memory costs, but include allocation costs on each multigrid level by the coarsening factor 1/8 in 3d, which results in the recursion factor $\sum_{\ell=0}^L 8^{\ell - L}$, where $L$ denotes the finest level. All variables are stored as doubles (8 Byte) and thus, the theoretically required memory is given by  
\begin{equation} \label{eq:memory}
m_t = 3\,(n_{L,u} + n_{L,p}) \left( \sum_{\ell = 0}^L 8^{\ell-L} \right)  8~ \text{(Byte)}.
\end{equation}
For level $L=7$, the theoretical memory consumption is then $m_t = 13.63 \,\text{(GB)}$ and the actual measured memory consumption $m_a=13.87 \,\text{(GB)}$, which differs by less than 1.7\%. Summarizing, the UMG method uses less than 44\% of the actual consumable memory and can perform computations with up to $5.3 \cdot 10^8$ DoFs on a conventional low cost workstation in about 12 minutes. 
On level $L = 7$, the SCG algorithm has a theoretical memory consumption of $m_t = 25.95\, \text{(GB)}$ (nearly twice the memory of the UMG method) and the PMINRES method would require $m_t = 51.90\, \text{(GB)}$ of memory, which is not possible. Note, the available 
memory for the computation is  7.7\% less than total memory, since 
system processes also occupy memory. 

Additionally, we study the different solvers for the Stokes system for the $D$-operator formulation. Here, the main difference is the $A_2$-block, which consists now of 9 sparse blocks, instead of 3 diagonal blocks for the Laplace-operator. The memory consumption for the $D$-operator formulation is almost the same as in the case of the Laplace-operator, which is due to the matrix-free implementation and the block structure.
Thus only a very minor part of the memory is needed to allocate the operators.

\begin{table}[h!]
\centering
\begin{tabular}{| l  r | r  r  r | r  r  r | r  r  r |}
\hline
\multicolumn{2}{|c|}{}  & \multicolumn{3}{c|}{SCG} & \multicolumn{3}{c|}{PMINRES} & \multicolumn{3}{c|}{UMG}\\
\hline
$L$ & DoFs & iter & \multicolumn{1}{c}{time} & ratio & iter & \multicolumn{1}{c}{time} & ratio & iter & \multicolumn{1}{c}{time} & ratio \\
\hline
2 &  $1.5 \cdot 10^{4\ }$	& 18	&     0.14	& 1.25	& 77	&     0.25	& 1.07	&   9 &     0.14	& 1.78 \\ 
3 &  $1.3 \cdot 10^{5\ }$	& 17	&     0.79	& 1.35	& 74	&     1.20	& 1.53	&   8 &     0.52	& 1.81 \\ 
4 &  $1.0 \cdot 10^{6\ }$	& 16	&     5.91	& 1.69	& 68	&     8.09	& 2.00	&   8 &     3.49	& 1.95 \\ 
5 &  $8.2 \cdot 10^{6\ }$	& 16	&   44.43	& 1.74	& 65	&   56.01	& 2.08	&   8 &   26.61		& 2.10\\
6 &  $6.6 \cdot 10^{7\ }$	& 16	& 351.89	& 1.64 	& 62	& 416.96	& 2.17	&   8 & 208.65 		& 2.18\\
7 & $5.3 \cdot 10^{8\ }$	& 16 & 2559.75 & 1.67 & \multicolumn{3}{c|}{out of memory} & 8 & 1645.86 & 2.25 \\
\hline
\end{tabular}
\caption{Iteration numbers and time-to-solution (in sec.) for the $D$-operator and ratios of the time-to-solution for Laplace- and $D$-operator.}
\label{T:itertime_D_ser}
\end{table}

In Tab.~\ref{T:itertime_D_ser}, we present the corresponding results in form of iteration numbers and time-to-solution. Again, we obtain robustness with respect to the refinement level for all considered solvers and observe that the UMG method is also in this case the most efficient solver. Note, that the difference of the SCG algorithm and UMG method is now roughly a factor 1.7 with respect to the time-to-solution. Also, the SCG algorithm profits in the case of the $D$-operator in terms of iteration numbers. The reason 
is that this operator directly couples the velocity components $\mathbf{u}_i$, $i=1,\dots,3$ and thus information can be transported more efficiently, while in case of the Laplace-operator the information is exclusively exchanged via the $B$-block.


For a comparison of the Laplace- and $D$-operator formulation, we also include the ratios of the time-to-solution for the two considered operators in Tab.~\ref{T:itertime_D_ser}, for the serial test cases.  While this factor on level $L=6$ is approximately 1.6 for the SCG algorithm, it is roughly 2.2 for the PMINRES method and the UMG method. Note, the ratios are increasing, since the difference in the number of non-zero entries (with respect to both formulations) increases with increasing problem size and thus a single core has to perform more flops.

\subsection{Operator and performance evaluations}

Further, we study the number of operator evaluations for the individual blocks and solvers. Let us denote by $n_{\text{op}, \ell}(\cdot) \in \mathbb{N}$ the number of operator evaluations on level $\ell$ for one of the considered blocks, then the total number of evaluations per block, weighted according to the workload, is determined by
\begin{align}\label{op_count}
n_\text{op}(\cdot) = \sum_{\ell=0}^L 8^{\ell - L} n_{\text{op}, \ell}(\cdot).
\end{align}
Note, operator evaluations are performed during the smoothing process, matrix-vector multiplication and the residual calculation. In the following, we exemplarily describe the derivation for the UMG method, using a variable $V$--cycle, \Vvar$(3,3)$. Within one smoothing step of the inexact Uzawa method, cf. \eqref{Uzawa_smooth}, we have two applications of $A_i$ (SHGS, where the part of the residual is included in the smoother), two applications of $B$ and one application of $C$. This results for level $L=6$ with $n_I = 8$ iterations, cf. Tab.~\ref{T:itertime1}, in
\begin{align*}
n_\text{op}(A) &= n_I \sum_{\ell=0}^L 8^{\ell-L}  (4(3 + 2(L-\ell)) + 1) \approx 129.31,\\
n_\text{op}(B) &= n_I \sum_{\ell=0}^L 8^{\ell-L}  (4(3 + 2(L-\ell)) + 2) \approx 138.45.
\end{align*}
Note, $n_\text{op}(C) = n_\text{op}(B)/2 \approx 69.22$. Further, $n_\text{op}(M)$ is quite small since it is only applied on the coarse grid, in case of the UMG method.

\definecolor{darkgreen}{rgb}{0.125,0.5,0.169}
\begin{figure}[h!]
\centering
\begin{tikzpicture}[scale=0.8]
  \begin{axis}[
    ybar,
    enlarge x limits=0.25,
    legend style={at={(0.7475,1.001)},
    anchor=north,legend columns=-1},
    ylabel={\# operator evaluations},
    symbolic x coords={SCG, PMINRES, UMG},
    xtick=data,
    nodes near coords,
    every node near coord/.append style={font=\scriptsize, inner sep=2pt, opacity=1},
    ]
    \addplot[color = blue, fill = blue, fill opacity=0.5] coordinates {(SCG,463) (PMINRES,305) (UMG,129)};
    \addplot[color = red, fill = red, fill opacity=0.5] coordinates {(SCG,72) (PMINRES,136) (UMG,138)};
    \addplot[color = darkgreen, fill = darkgreen, fill opacity=0.5] coordinates {(SCG,36) (PMINRES,68) (UMG,69)};
    \addplot[color = gray, fill = gray, fill opacity=0.5] coordinates {(SCG,31) (PMINRES,68) (UMG, 0)};
    \legend{$A_1$, $B$, $C$, $M$}
  \end{axis}
\end{tikzpicture}
\hspace*{2.5em}
\begin{tikzpicture}[scale=0.8]
  \begin{axis}[
    ybar,
    ymax=507,
    enlarge x limits=0.25,
    legend style={at={(0.7475,1.001)},
    anchor=north,legend columns=-1},
    ylabel={\# operator evaluations},
    symbolic x coords={SCG, PMINRES, UMG},
    xtick=data,
    nodes near coords,
    every node near coord/.append style={font=\footnotesize, inner sep=2pt, opacity=1},
    ]
    \addplot[color = blue, fill = blue, fill opacity=0.5] coordinates {(SCG, 273) (PMINRES, 282) (UMG,129)};
    \addplot[color = red, fill = red, fill opacity=0.5] coordinates {(SCG, 44) (PMINRES, 126) (UMG, 138)};
    \addplot[color = darkgreen, fill = darkgreen, fill opacity=0.5] coordinates {(SCG, 22) (PMINRES, 63) (UMG, 69)};
    \addplot[color = gray, fill = gray, fill opacity=0.5] coordinates {(SCG, 16) (PMINRES, 63) (UMG, 0)};
    \legend{$A_2$, $B$, $C$, $M$}
  \end{axis}
\end{tikzpicture}
\caption{Number of operator evaluations $\lfloor n_\text{op} \rceil$ for the Laplace- (left) and $D$-operator (right) on level $L = 6$, where $\lfloor \cdot \rceil$ denotes the rounding to nearest integer.}
\label{F:opcount1}
\end{figure}
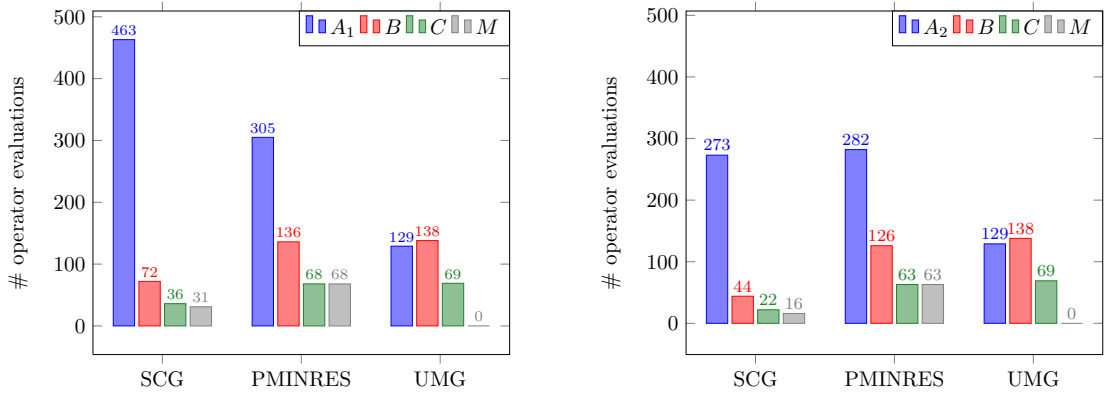

The results for refinement level $L = 6$ are presented in Fig.~\ref{F:opcount1}. Let us discuss the case of the Laplace-operator first. While the number of operator evaluations for $B$ and $C$ are similar for the PMINRES- and UMG method, and half the size for the SCG algorithm, the evaluations for the $A_1$ block differ significantly. Here, we observe that the UMG method has less than 50\% of the evaluations of the PMINRES method and less than 30\% compared to the SCG algorithm. The UMG method profits significantly from the lower 
operator count for $A_1$, which is also reflected by the time-to-solution, cf. Tab.~\ref{T:itertime1}. Also, we present the number of operator evaluations for the case of the $D$-operator in the right of Fig.~\ref{F:opcount1} for refinement level $L = 6$. The number of operator evaluations are comparable to the Laplace case (scaled by a factor), where fewer applications occur in case of the SCG algorithm and the PMINRES method, although this is not so significant in the latter case. Note, the number of operator evaluations in case of the UMG method stays exactly the same. This behavior is also reflected by the time-to-solution, cf. Tab.~\ref{T:itertime_D_ser}.

For the comparison of Laplace- and $D$-operator formulation, we include the ratios using the number of operator evaluations. For both formulations, we define the total number of operator evaluations for both formulations by
\begin{align*}
n_\text{total}^\text{Lap.} &= 3\, n_\text{op}^\text{Lap.}(A) +  3\, n_\text{op}^\text{Lap.}(B) + n_\text{op}^\text{Lap.}(C) + n_\text{op}^\text{Lap.}(M),\\
n_\text{total}^D &= 3\, \mu_\text{d}\, n_\text{op}^D(A) +  3\, n_\text{op}^D(B) + n_\text{op}^D(C) + n_\text{op}^D(M),
\end{align*}
where $n_\text{op}^\text{Lap.}(\cdot)$ and $n_\text{op}^D(\cdot)$ denotes the number of operator evaluations for the Laplace- and $D$-operator formulation. Moreover, $\mu_\text{d} = 3.25$ denotes the measured multiplicative factor in the application of one FHGS smoothing step for $A_1$ and $A_2$ with respect to time.

\begin{table}[h!]
\centering
\begin{tabular}{ l | p{9em} p{9em} p{9em} }
\hline
 & \multicolumn{1}{c}{SCG} &  \multicolumn{1}{c}{PMINRES} &  \multicolumn{1}{c}{UMG}\\
\hline
 $n_\text{total}^D / n_\text{total}^\text{Lap.}$	&  \multicolumn{1}{c}{1.69}	&  \multicolumn{1}{c}{2.23}	&  \multicolumn{1}{c}{2.00}\\
\hline
\end{tabular}
\caption{Theoretical expected ratios with respect to the total number of operator evaluations on level $L=6$.}
\label{T:ratios_opcoutns}
\end{table}

In Tab.~\ref{T:ratios_opcoutns}, we present the ratios of the total number of operator evaluations, which are in good agreement with those calculated from the time-to-solution, cf. Tab.~\ref{T:itertime_D_ser}, with differences of less than 10\%.

\subsection{Parallel performance}\label{sec:par_perfrom}
In the following, we present numerical results for the massively parallel case. Beside the comparison of the different solvers, also the different formulations of the Stokes systems are studied. Scaling results are present for up to 786\,432 threads and $10^{13}$ degrees of freedom. We consider 32 threads per node and 2 tetrahedrons of the initial mesh $\mathcal{T}_{-2}$ per thread. The finest computational level is obtained by a 7 times refined initial mesh. If not further specified, the settings for the solvers remain as described above.

\begin{table}[h!]
\centering
\begin{tabular}{|r r r | l r | l r | l r |}
        \hline
        \multicolumn{3}{|c|}{} & \multicolumn{2}{c|}{SCG} & \multicolumn{2}{c|}{PMINRES} & \multicolumn{2}{c|}{UMG}  \\ \hline
	\multicolumn{1}{|c}{nodes} 	& \multicolumn{1}{c}{threads}	& \multicolumn{1}{c|}{DoFs}	& iter		& time 	& iter 	& time	& iter	& time  \\ \hline
	1		& 24			& $6.6 \cdot 10^{7\ }$	& 28 (3)   & 137.26 	& 65 (3)   & 116.35	& 7  	&   62.44 \\ 
	6		& 192		& $5.3 \cdot 10^{8\ }$	& 26 (5)   & 135.85 	& 64 (7)   & 133.12	& 7 	&   70.97 \\ 
	48		& 1\,536		& $4.3 \cdot 10^{9\ }$	& 26 (15) & 139.64 	& 62 (15) & 133.29	& 7 	&   79.14 \\ 
	384		& 12\,288		& $3.4 \cdot 10^{10}$	& 26 (30) & 142.92 	& 62 (35) & 138.26	& 7 	&   92.76 \\ 
    3\,072		& 98\,304		& $2.7 \cdot 10^{11}$	& 26 (60) & 154.73 	& 62 (71) & 156.86	& 7 	& 116.44 \\
  24\,576		& 786\,432	& $2.2 \cdot 10^{12}$	& 28 (120) & 187.42 	& 64 (144) & 184.22		& 7	& 169.15 \\
	\hline
\end{tabular}
\caption{Weak scaling results with time-to-solution (in sec.) for the Laplace-operator.}
\label{T:itertime_lap_par}
\end{table}

\begin{table}[h!]
\centering
\begin{tabular}{|r r r | l r | l r | l r |}
        \hline
        \multicolumn{3}{|c|}{} & \multicolumn{2}{c|}{SCG} & \multicolumn{2}{c|}{PMINRES} & \multicolumn{2}{c|}{UMG}  \\ \hline
	\multicolumn{1}{|c}{nodes} 	& \multicolumn{1}{c}{threads}	& \multicolumn{1}{c|}{DoFs}	& iter		& time 	& iter 	& time	& iter 	& time  \\ \hline
	1		& 24			& $6.6 \cdot 10^{7\ }$	& 28 & 136.30	& 65 & 115.05 & 7 & 58.80 \\ 
	6		& 192		& $5.3 \cdot 10^{8\ }$	& 26 & 134.26 	& 64 & 130.56 & 7 & 64.40 \\ 
	48		& 1\,536		& $4.3 \cdot 10^{9\ }$	& 26 & 135.06	& 62 & 128.34 & 7 & 65.03 \\ 
	384		& 12\,288		& $3.4 \cdot 10^{10}$	& 26 & 135.41	& 62 & 128.34 & 7 & 64.96 \\ 
    3\,072		& 98\,304		& $2.7 \cdot 10^{11}$	& 26 & 139.55 	& 62 & 133.30 & 7 & 66.08 \\
  24\,576		& 786\,432	& $2.2 \cdot 10^{12}$	& 28 & 154.06 	& 64 & 139.52 & 7 & 68.46 \\
	\hline
\end{tabular}
\caption{Weak scaling results with time-to-solution (in sec.) without coarse grid time for the Laplace-operator.}
\label{T:itertime_lap_withoutcoarse_par}
\end{table}

In Tab.~\ref{T:itertime_lap_par}, we present weak scaling results for the individual solvers in the case of the Laplace-operator. Additionally, we present the coarse grid iteration numbers in brackets, which in case of the SCG- and MINRES solver represent the CG iterations, cf. Sec.~\ref{Sec:coarse_grid}. We observe excellent scalability of all three solvers, where the SCG algorithm and PMINRES method reach similar time-to-solution for the considered test. The UMG method is again faster compared to the other solvers, but a factor of two as in the serial test case is not reached, cf. Tab.~\ref{T:itertime1}. To get a better understanding, we also present weak scaling results in
Tab.~\ref{T:itertime_lap_withoutcoarse_par} where the time of the coarse grid is excluded.
Here, we again observe the factor of two in the time-to-solution in favor of the UMG method.

\begin{table}[h!]
\centering
\begin{tabular}{|r r r | l r | l r | l r |}
        \hline
        \multicolumn{3}{|c|}{} & \multicolumn{2}{c|}{SCG} & \multicolumn{2}{c|}{PMINRES} & \multicolumn{2}{c|}{UMG}  \\ \hline
	\multicolumn{1}{|c}{nodes} 	& \multicolumn{1}{c}{threads}	& \multicolumn{1}{c|}{DoFs}	& iter		& time 	& iter 	& time	& iter	& time  \\ \hline
	1		& 24			& $6.6 \cdot 10^{7\ }$ &	  16 (5) & 177.10 & 58 (5) & 191.03		& 7	& 101.34 \\ 
	6		& 192		& $5.3 \cdot 10^{8\ }$ &	16 (10) & 186.29 & 56 (10) & 204.89		& 7	& 112.85 \\ 
	48		& 1\,536		& $4.3 \cdot 10^{9\ }$ &	16 (20) & 191.51 & 56 (20) & 211.00		& 7	& 122.16 \\ 
	384		& 12\,288		& $3.4 \cdot 10^{10}$ &	16 (40) & 197.14 & 54 (40) & 208.86		& 7	& 142.51 \\ 
    3\,072		& 98\,304		& $2.7 \cdot 10^{11}$ &	16 (80) & 209.30 & 54 (80) & 227.97				& 7	& 164.72 \\
  24\,576		& 786\,432	& $2.2 \cdot 10^{12}$ &   16 (160) & 226.33 & 54 (160)	& 255.25		& 7	& 217.98 \\
	\hline
\end{tabular}
\caption{Weak scaling results with time-to-solution (in sec.) for the $D$-operator.}
\label{T:itertime_D_par}
\end{table}

\begin{table}[h!]
\centering
\begin{tabular}{|r r r | l r | l r | l r |}
        \hline
        \multicolumn{3}{|c|}{} & \multicolumn{2}{c|}{SCG} & \multicolumn{2}{c|}{PMINRES} & \multicolumn{2}{c|}{UMG}  \\ \hline
	\multicolumn{1}{|c}{nodes} 	& \multicolumn{1}{c}{threads}	& \multicolumn{1}{c|}{DoFs}	& iter		& time 	& iter 	& time	& iter 	& time  \\ \hline
	1		& 24			& $6.6 \cdot 10^{7\ }$	& 16 & 176.51  	& 58 & 188.58 & 7 &  96.67 \\ 
	6		& 192		& $5.3 \cdot 10^{8\ }$	& 16 & 183.85 	& 56 & 203.20 & 7 & 102.18 \\ 
	48		& 1\,536		& $4.3 \cdot 10^{9\ }$	& 16 & 187.82   & 56 & 204.66 & 7 & 104.50 \\ 
	384		& 12\,288		& $3.4 \cdot 10^{10}$	& 16 & 189.50   & 54 & 197.27 & 7 & 105.87 \\ 
    3\,072		& 98\,304		& $2.7 \cdot 10^{11}$	& 16 & 193.51   & 54 & 205.13 & 7 & 107.20 \\ 
  24\,576		& 786\,432	& $2.2 \cdot 10^{12}$	& 16 &  203.77 	& 54    & 205.25       & 7 & 100.04 \\ 
	\hline
\end{tabular}
\caption{Weak scaling results with time-to-solution (in sec.) without coarse grid time for the $D$-operator.}
\label{T:itertime_D_withoutcoarse_par}
\end{table}

We also include numerical results for the $D$-operator, which are presented in Tab.~\ref{T:itertime_D_par} and Tab.~\ref{T:itertime_D_withoutcoarse_par}. Beside robustness with respect to the problem size of all solvers, we observe that the iteration numbers are smaller for the same relative accuracy in comparison to the Laplace-operator, for the SCG algorithm and the PMINRES method, which is due to the cross coupling terms in the $A_2$-block. In case of the UMG method, the numbers are the same as in the Laplace-operator case. Perfect scalability is observed for the multilevel part, cf. Tab.~\ref{T:itertime_D_withoutcoarse_par}. Note, the coarse grid times are of similar order as in the Laplace-operator case. 

\begin{table}[h!]
\centering
\begin{tabular}{ | r | r | r | r |}
\hline
DoFs & \multicolumn{1}{>{\centering}p{4.5em}|}{SCG} & \multicolumn{1}{>{\centering}p{4.5em}|}{PMINRES} & \multicolumn{1}{>{\centering}p{4.5em}|}{UMG} \\
\hline
$6.6 \cdot 10^{7\ }$	& 1.29 & 1.64 & 1.62 \\ 
$5.3 \cdot 10^{8\ }$	& 1.37 & 1.54 & 1.59 \\ 
$4.3 \cdot 10^{9\ }$	& 1.37 & 1.58 & 1.54 \\ 
$3.4 \cdot 10^{10}$	& 1.38 & 1.51 & 1.54 \\
$2.7 \cdot 10^{11}$	& 1.35 & 1.45 & 1.41 \\
$2.2 \cdot 10^{12}$	& 1.20 & 1.39 & 1.29 \\
\hline
\end{tabular}
\caption{Ratios of the time-to-solution for Laplace- and $D$-operator, the parallel case.}
\label{T:factor_times_par}
\end{table}

For a better comparison of the Laplace- and $D$-operator results, we include Tab.~\ref{T:factor_times_par}, representing the ratios of the time-to-solution. These numbers tend to be comparable, for the different solvers. In the weak scaling set-up each thread performs the same amount of work and thus the ratios stay almost constant for increasing problem sizes, whereas in the serial case increasing ratios were observed, cf. Tab.~\ref{T:itertime_D_ser}. 
Furthermore, the communication costs are identical for both formulations.
However, the workload for both formulations differs for a single process.

\subsection{Exploring the DoF limits} 
In this section, we push the number of degrees of freedom to the limit. The main limitation, we are facing, is the total memory of the computing system.
One compute node on JUQUEEN owns 16 GB of shared memory.
In order to maximize the memory that can be used on each node,
and the number of unknowns of one thread with a uniform workload balance, 
we reduce the number of threads per node to 16 and assign 3 tetrahedrons of the initial mesh per thread. By reducing the total number of tetrahedrons on a node, a refinement up to level 8 of the initial grid is possible. The UMG method is employed,
since the other solvers require more memory.
As a computational domain, we apply the spherical shell
$\Omega = \{ \mathbf{x} \in \mathbb{R}^3 : 0.55 < \| \mathbf{x} \|_2 < 1 \}$.
This geometry could typically appear in simulations for molecules,
quantum mechanics, or geophysics.
The initial mesh $\mathcal{T}_{-2}$ consists of 240 tetrahedrons for the case of 5 nodes and 80 threads. The number of degrees of freedoms on the coarse grid
$\mathcal{T}_0$ grows from $9.0 \cdot 10^3$ to $4.1 \cdot 10^7$
by the weak scaling.
We consider the Stokes system with the Laplace-operator formulation.
The relative accuracies for coarse grid solver (PMINRES and CG algorithm) are set to
$10^{-3}$ and $10^{-4}$, respectively.
All other parameters for the solver remain as previously described.
\begin{table}[h]
\centering
\begin{tabular}{|r r l | l r r r |}
        \hline
	\multicolumn{1}{|c}{nodes} 	& \multicolumn{1}{c}{threads}	& \multicolumn{1}{c|}{DoFs}	& iter		& time & time w.c.g. & time c.g. in \% \\ \hline
	   5	& 80			& $2.7 \cdot 10^{9} $		& 10 & 685.88 & 678.77	& 1.04 \\ 
	 40	& 640		& $2.1 \cdot 10^{10}$ 	& 10 & 703.69 & 686.24	& 2.48 \\ 
	320	& 5\,120		& $1.2 \cdot 10^{11}$	& 10 & 741.86 & 709.88	& 4.31 \\ 
    2\,560	& 40\,960		& $1.7 \cdot 10^{12}$ 	&   9 & 720.24 & 671.63	& 6.75 \\
  20\,480	& 327\,680	& $1.1 \cdot 10^{13}$	&   9 & 776.09 & 681.91	& 12.14 \\
	\hline
\end{tabular}
\caption{Weak scaling results with and without coarse grid for the spherical shell geometry.}
\label{T:itertime_lap_sp_par}
\end{table}

Numerical results with up to $10^{13}$ degrees of freedom are presented in
Tab.~\ref{T:itertime_lap_sp_par}, where we observe robustness with respect to the problem size and excellent scalability. Beside the time-to-solution (time) we also present the time excluding the time necessary for the coarse grid (time w.c.g.) and the total amount in \% that is needed to solve the coarse grid. For this particular setup, this fraction does not exceed 12\%. 
Due to 8 refinement levels, instead of 7 previously, and the reduction of threads per node
from 32 to 16, longer computation times (time-to-solution) are expected, compared to the results in Sec.~\ref{sec:par_perfrom}. In order to evaluate the performance, we compute the factor $t \, n_c\, n^{-1}$, where $t$ denotes the time-to-solution (including the coarse grid), $n_c$
the number of used threads, and $n$ the degrees of freedom.
This factor is a measure for the compute time per degree of freedom, weighted with the number of threads, under the assumption of perfect scalability. 
For $1.1 \cdot 10^{13}$ DoFs, this factor takes the value of
approx. $2.3 \cdot 10^{-5}$ and for the case of $2.2 \cdot 10^{12}$ DoFs on the unit cube (Tab.~\ref{T:itertime_lap_par}) approx. $6.0 \cdot 10^{-5}$, which is of the same order.
Thus, in both scaling experiments the time-to-solution for one DoF is comparable.
The reason why the ratio is even smaller for the extreme case of $1.1 \cdot 10^{13}$
DoFs is the deeper multilevel hierarchy.
Recall also that the computational domain is different in both cases.

The computation of $10^{13}$ degrees of freedom is close to the limits that are given by
the shared memory of each node. By \eqref{eq:memory}, we obtain a theoretical total memory consumption of 274.22 TB, and on one node of 14.72 GB. 
Though 16 GB of shared memory per node is available, 
we employ one further optimization step and do not allocate the right-hand side on the finest grid level. The right-hand side vector is replaced by an assembly on-the-fly, i.e., 
the right-hand side values are evaluated and integrated locally when needed.
By applying this on-the-fly assembly, the theoretical memory consumption (without 
storage of the finest level right-hand side) becomes
\begin{equation*}
m_t = 2\,(n_{L,u} + n_{L,p})\,8  +  3\,(n_{L,u} + n_{L,p}) \left( \sum_{\ell = 0}^{L-1} 8^{\ell-L} \right) 8~ \text{(Byte)}.
\end{equation*}
This yields for $L = 6$ a total memory of $m_t = 198.24\, \text{(TB)}$ and $m_t = 9.71\, \text{(GB)}$ per node, respectively. 
As discussed in the serial test case, see Sec.~\ref{timetosol}, the theoretical memory consumption $m_t$ is in very good agreement with the actual memory consumption $m_a$ and thus, just results for $m_t$ are presented here.
Note, this memory consumption corresponds to 60.70\% of the entire
available memory on the 20\,480 nodes that are employed.

We recall here that the matrix-free implementation and the block-structured approach of HHG 
are essential to reach the presented performance levels, since otherwise the storage of the assembled matrix
$\mathcal{K}$ would  by far become the dominating contribution to the memory requirements.
Therefore only much smaller systems could be solved by conventional iterative methods.

\section{Textbook multigrid efficiency}\label{sec:tme} 
Due to the architectural complexity of modern computers, a systematic performance analysis is essential for an appropriate design of scientific computing software for high performance computers.  
In this section, we apply the idea of {\em textbook multigrid efficiency} (TME) as performance analysis technique,
see, e.g., \cite{brandt1998barriers,gmeiner-ruede-stengel-waluga-wohlmuth_2015_2}, to the UMG method for the Laplace-operator formulation of the Stokes system.  
The original TME metric quantifies 
the computational cost of a solver in a problem specific basic work unit (WU).
Originally, the TME metric \cite{brandt1998barriers} postulates
a cost of less than 10 WU for a well-designed multigrid method.
In \cite{gmeiner-ruede-stengel-waluga-wohlmuth_2015_2} an extension to {\em parallel textbook multigrid efficiency} (parTME) is proposed by interlinking the classical algorithmic TME efficiency 
with efficient scalable implementations to
\begin{equation}
 E_\text{parTME} \coloneqq \frac{t(n, n_c)}{t_\text{\tiny WU}(n, n_c)} = \frac{t(n, n_c)\, n_c\, \mu_\text{sm}}{n}.
\label{equ:partme}
\end{equation}
Here, $t(n, n_c)$ denotes the time-to-solution necessary for $n$ unknowns computed by $n_c$ threads and $t_\text{\tiny WU}(n, n_c)$ is the time for one work unit. 
When the performance of a single thread is given by $\mu_{\text{sm}}$ lattice updates per second (Lups/s) such that $n_c$ perfectly  scalable threads would perform $n_c \mu_{\text{sm}}\,\text{Lups/s}$. Clearly, the parTME efficiency metric in \eqref{equ:partme} depends on the underlining computing system and the predicted and actually measured \text{Lups/s}. 
A deeper analysis can be based  on enhanced node performance models like the one 
proposed in \cite{gmeiner-ruede-stengel-waluga-wohlmuth_2015_2}.

In order to achieve full textbook efficiency, we present results for the UMG method employed
with a full multigrid (FMG) algorithm \cite{adams_2014,brandt1982guide,hackbusch_1985,trottenberg-oosterlee-schuller_2000}. Multigrid methods achieve level independent convergence rates with optimal complexity. Nevertheless, lower discretization errors on finer levels require 
additionally better approximations of the algebraic system on finer levels.
Only the FMG algorithm yields an optimal complexity solver in this context
since $V$-- and $W$--cycles alone would require a higher number of  iterations
on finer levels since they must asymptotically be driven to increasingly lower stopping tolerances.

Therefore, the FMG algorithm is the natural candidate to study and evaluate the TME and parTME efficiency. We define one work unit (WU) by the computational cost of one discrete Stokes system application, which consists of 10 blocks (3 blocks for $A_1$, 6 blocks for $B$, $B^\top$ and 1 block  for $C$). Note that one smoothing step of the UMG method including the residual evaluation consists of applying 2 times $A_1$ (6 blocks), 2 times $B$ (6 blocks) and once $C$ (1 block), which results together in $13 / 10$ WU.

In the following, we choose as computational domain $\Omega = (0,1)^3$ and, as in \cite{gmeiner-ruede-stengel-waluga-wohlmuth_2015_2}, the analytical solution
\begin{align*}
\mathbf{u} = 
\begin{pmatrix}
-4 \cos(4 x_3)\\
8 \cos(8 x_1)\\
-2 \cos (2 x_2)
\end{pmatrix},
\qquad p = \tilde{p} - |\Omega|^{-1} \langle \tilde{p}, 1 \rangle_\Omega, \quad \tilde{p} = \sin(4 x_1) \sin(8 x_2) \sin (2 x_3).
\end{align*}
The right-hand side and the Dirichlet boundary values are chosen accordingly. Then for a discrete solution vector $(\mathbf{u}, \mathbf{p})^\top \in \mathbb{R}^{n_{\ell,u} + n_{\ell,p}}$ with components for velocity and pressure, we define the mesh dependent norm
\begin{align*}
\| (\mathbf{u}, \mathbf{p})^\top \|_h \coloneqq \left( \| \mathbf{u} \|_2^2 + h_\ell^2\ \| \mathbf{p} \|_2^2 \right)^{1/2},
\end{align*}
as commonly used, see, e.g., \cite{larin-reusken_2008}, where $\| \cdot \|_2$ denotes the Euclidean norm. 

Before, the computational cost of the FMG algorithm is analyzed, we address the accuracy achieved by different FMG algorithm variants. Therefore, we introduce the ratio
\begin{align*}
\gamma_\ell \coloneqq  \frac{\| (\mathbf{I}_\ell(\mathbf{u}) - \tilde{\mathbf{u}}_\ell, \mathbf{I}_\ell(p) - \tilde{\mathbf{p}}_\ell)^\top \|_h}{\| (\mathbf{I}_\ell(\mathbf{u}) - \mathbf{u}_\ell, \mathbf{I}_\ell(p) - \mathbf{p}_\ell)^\top \|_h},
\end{align*}
which is an indicator for the deviation of the total error $\| (\mathbf{I}_\ell(\mathbf{u}) - \tilde{\mathbf{u}}_\ell, \mathbf{I}_\ell(p) - \tilde{\mathbf{p}}_\ell)^\top \|_h$ (discretization- and algebraic error) and the discretization error $\| (\mathbf{I}_\ell(\mathbf{u}) - \mathbf{u}_\ell, \mathbf{I}_\ell(p) - \mathbf{p}_\ell)^\top \|_h$ with respect to the nodal interpolation vector $(\mathbf{I}_\ell(\mathbf{u}), \mathbf{I}_\ell(p))^\top$ of the analytic solution for velocity and pressure. We are interested in a $\gamma_\ell$ close to one, since it indicates that the finite element discretization error is about as large as the total error which includes the remaining algebraic error
of the computed solution. 

\begin{figure}[h!]
\centering
\begin{tikzpicture}[scale= 1]
\begin{axis}[
	axis lines=none,
	width=\textwidth,
	height=0.25\textwidth,
	]
	\addplot[color=black, mark=*] coordinates {
	(0,0) (1,1) (2,0) (3,1) (4,2) (5,1) (6,0) (7,1) (8,2)
	};
	\addplot[color=black, mark=*] coordinates {
	(15,0) (16,1) (17,0) (18,1) (19,0) (20,1) (21,2) (22,1) (23,0) (24,1) (25,2) (26,1) (27,0) (28,1) (29,2)
	};
\end{axis}
\end{tikzpicture}
\caption{Illustration of a full multigrid (FMG) V-cycle and 2V-cycle.}
\label{F:FMG}
\end{figure}
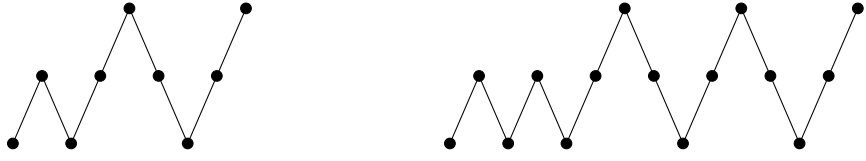

In our numerical examples, two variants of the FMG algorithm are considered with one and two variable $V$--cycle per multigrid level, respectively, as illustrated in Fig.~\ref{F:FMG}. Additionally, we vary the number of pre- and post-smoothing steps. We start with a zero initial vector for velocity and pressure. 
%
%
\begin{table}[h!]
\centering
\setlength{\tabcolsep}{0.5em}
\begin{tabular}{| l | l l l l l l l l |}
        \hline
      DoFs		 &  \Vvar(1,1)	& 2\Vvar(1,1)	&	\Vvar(2,2)	&	2\Vvar(2,2)	& \Vvar(3,3)	& 2\Vvar(3,3)	& \Vvar(5,5)	& 2\Vvar(5,5)  \\ \hline
$8.0\cdot 10^6$		&	4.4152	&	2.1132	&	3.4758	&	1.0752	&	1.4925	&	0.9955	&	2.0057	&      0.9969 \\
$6.8\cdot10^7$        	&	1.9826	&	1.0348	&	1.2035	&	1.0000	&	1.2173  &	1.0004  &	1.2002	&      1.0004 \\
$5.2\cdot10^8$        	&	1.4810	&	1.0067	&	1.1994	&	1.0013	&	1.1148  &	1.0004  &	1.1786	&      0.9990 \\
$4.4\cdot10^9$		&	1.4841	&	1.0056	&	1.1066	&	0.9989	&	1.1642	&	0.9990	&	1.1716	&      0.9974 \\
$3.4\cdot10^{10}$	&	1.5366	&	1.0075	&	1.2299	&	1.0001	&	1.2166	&	0.9967	&	1.1948	&      0.9941 \\
	\hline
\end{tabular}
\caption{Ratios $\gamma_\ell$ for different FMG variants.}
\label{T:gamma_ratio}
\end{table}

In Tab.~\ref{T:gamma_ratio}, we present the ratios $\gamma_\ell$ for different variants of the FMG algorithm, where the same coarse grid with 4\,096 tetrahedrons is considered for increasing refinement levels $\ell = 1, \dots, 5$, i.e., in the first row there is only one coarser level, and thus the results of a two-level algorithm are displayed. In each following row of the table, the mesh size is halved, and thus the number of multigrid level increases by one. The FMG algorithm with a \Vvar(2,2)--cycle shows that the total error is approximately 10\% - 20\% larger than the discretization error, except for the two-level algorithm which suffers from an inexactly computed correction on the coarse grid. 
Note, this effect completely vanishes with a larger level hierarchy. Similar to the observations in \cite{gmeiner-ruede-stengel-waluga-wohlmuth_2015_2}, the combined error can only be further reduced to the discretization error by the variant with two cycles on each level. 
The other variants considered in Tab.~\ref{T:gamma_ratio} are either more expensive (in work unit factors) despite no significant error reduction is achieved, or they are cheaper (in the work units) but then deliver an inadmissibly large error.
Thus, we choose for the the further TME and parTME efficiency considerations the \Vvar(2,2)--cycle within the FMG algorithm. Note, the values $\gamma_\ell$ are in some cases smaller than 1, which is due to small fluctuations.

A \Vvar(2,2)--cycle employs on level $\ell \in \mathbb{N}_0$ a total of
$2\, (2+2(L-\ell))\, 13 / 10$ WU, where $L \in \mathbb{N}$ denotes the finest level. Additionally, one WU has to be accounted for one residual computation per level. Further, we neglect the work for restriction and prolongation, but include the cost for the $V$--cycle recursion with standard coarsening in 3d by the factor $1/8$. Then, the FMG algorithm with a \Vvar(2,2)--cycle requires 
\begin{align*}
E_\text{TME} \coloneqq \lim_{L\rightarrow \infty}\sum_{k=0}^{L-1} 8^{-k} \sum_{\ell=1}^{L} 8^{\ell - L} \left(2\, (2 + 2 (L - \ell))\, \frac{13}{10} + 1\right) \approx 9~ \text{(WU)}.
\end{align*}
\begin{table}[h!]
\centering
\begin{tabular}{| r | l | l  l l l l l|}
	\hline
	& $E_\text{TME}$ 	& \multicolumn{6}{c|}{$E_\text{parTME}$} \\ \hline
DoFs	& 			&$8.2 \cdot 10^6$ & $6.6\cdot10^7$		& $5.3\cdot10^8$		& $4.3\cdot10^9$		& $3.4\cdot10^{10}$	& $2.7\cdot10^{11}$    	\\
$n_c$	& 			& 1 & 24				& 192				& 1\,536			& 12\,288		& 98\,304 		\\ \hline
FMG \Vvar(2,2) & 9.07	& 43.43 & 104.79		& 150.21		& 199.70		& 278.25	& 500.47 \\
FMG \Vvar(2,2) w.c.g. & 9.07		& 40.18 & 72.89				& 80.26				& 80.45				& 82.06			& 83.53			\\
	\hline
\end{tabular}
\caption{TME factors for different problem sizes and FMG cycles.}
\label{T:TME}
\end{table}

For the parallel TME factors, we consider a weak scaling with setup of the experiments as in Sec.~\ref{sec:par_perfrom}. Additionally, we include the (parallel) TME factors for one single thread. The efficiencies with and without coarse grid times (w.c.g.) are listed in Tab.~\ref{T:TME}. In \cite{gmeiner-ruede-stengel-waluga-wohlmuth_2015}, the parallel performance was studied on {JUQUEEN}. Since a simple memory bandwidth limit model was taken, the parallel measured and actual performance were mispredicted by a factor of 2. Only an enhanced model, as in \cite{gmeiner-ruede-stengel-waluga-wohlmuth_2015_2}, can reduce this mismatch to the actual performance.
Therefore, we employ the measured performance values for
$\mu_{\text{sm}} = 23.9\, \text{MLups/s}$.  We observe that the parTME efficiency
deteriorates significantly by up to a factor of 5 due to the suboptimal coarse grid solver and since coarser multigrid levels exhibit more overhead.
Since the classical TME efficiency assumes that the hierarchy of multigrid levels is large, 
respectively, infinity, the coarse grid cost is negligible there.
Hence for the parTME efficiency, we also present efficiency values without coarse grid. Here, a discrepancy of a factor of 4--9 between $E_\text{TME}$ and $E_\text{parTME}$ is observed. The reason therefore is that the classical $E_\text{TME}$ factor does not take into account the costs for restriction/prolongation and the communication. This discrepancy is also observed
for one single thread, cf. Tab.~\ref{T:TME}. The parTME efficiency stays almost constant in the transition to up to 98\,304 threads and $2.7\cdot10^{11}$ DoFs. Note, that TME and parTME could only coincide if there were 
no cost for intergrid transfer (restriction and prolongation), if the loops on the coarser grid were executed as efficiently as on the finest mesh, if there were no communication overhead, and if no additional copy-operations, or other program overhead occurred.



\section{Application to flow problems}\label{sec:appli} 
As an application of the proposed all-at-once multigrid method, we simulate an incompressible fluid flow around multiple spheres that are randomly placed in a cylindrical domain (sphere bed). Such problems appear in a wide range of applications, see, e.g., \cite{bogner-mohanty-ruede_2015, garcia_2011, dixon_2004, behr_2014}. 
One example is the simulation of fluid flow in chromatography columns, which is important for the purification of product molecules in biopharmaceutical industry involving the stationary Stokes system with up to $10^7$ DoFs  \cite{behr_2014}. We will present simulation results with up to $10^9$ DoFs. 
This model problem is a first step towards solving coupled multi-physics problems,
see, e.g., \cite{waluga-wohlmuth-ruede_2015}, using finite element discretizations
on massively parallel systems. Note, that beside the appropriate code design 
and developing a fast multigrid solver, 
also the visualization is challenging for such exceedingly sized simulation results.

\subsection{Problem description and boundary conditions}
\begin{figure}[h!]
\centering
\includegraphics[width=0.8\textwidth]{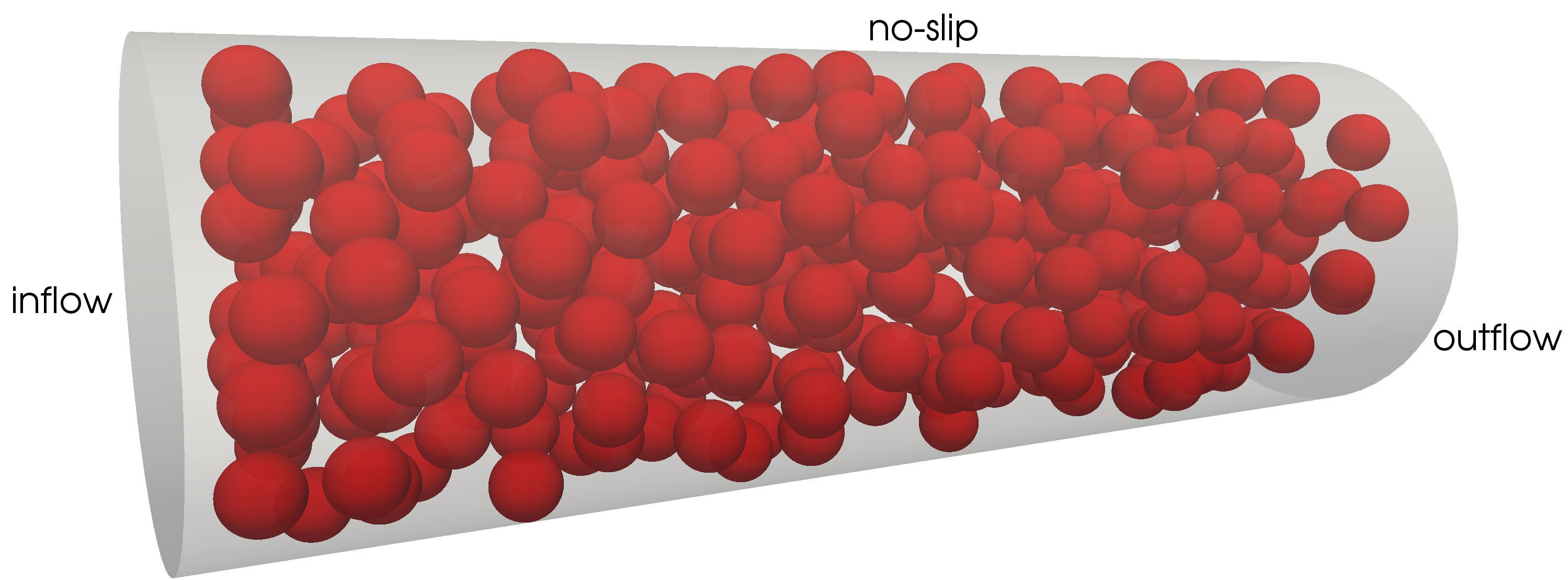}
\caption{Computational domain with $n_s = 305$ spheres with random distribution.}
\label{F:spheres_geom}
\end{figure}

In the following, we shall consider as a computational domain, similar to a so-called packed sphere bed, see, e.g., \cite{dixon_2004, behr_2014}. We consider a cylinder, aligned along the $x_1$-axis of length 6 and radius 1 with a given number, $n_s \in \mathbb{N}$, of spheres which are
obstacles, i.e., the domain is $\Omega = \{ \mathbf{x} \in \mathbb{R}^3\, :\, x_2^2 + x_3^2 < 1,\,  x_3 \in (0, 6) \} \setminus \overline{\mathcal{B}}$, where $\mathcal{B} = \bigcup_{i=1}^{n_s} B_r(\mathbf{c}_i)$ is the union of spheres with radius $r = 0.15$ 
that are located at the center points $\mathbf{c}_i \in \Omega$. The points $\mathbf{c}_i$ are randomly chosen, have a minimal distance of $2r$ from each other, and have at least a distance of $r$ from the cylinder boundary, see Fig.~\ref{F:spheres_geom}. Note, spheres might have a very small distance to each other or the cylinder boundary, which makes the problem even more challenging.
\begin{figure}[h!]
\centering
\includegraphics[width=0.4\textwidth]{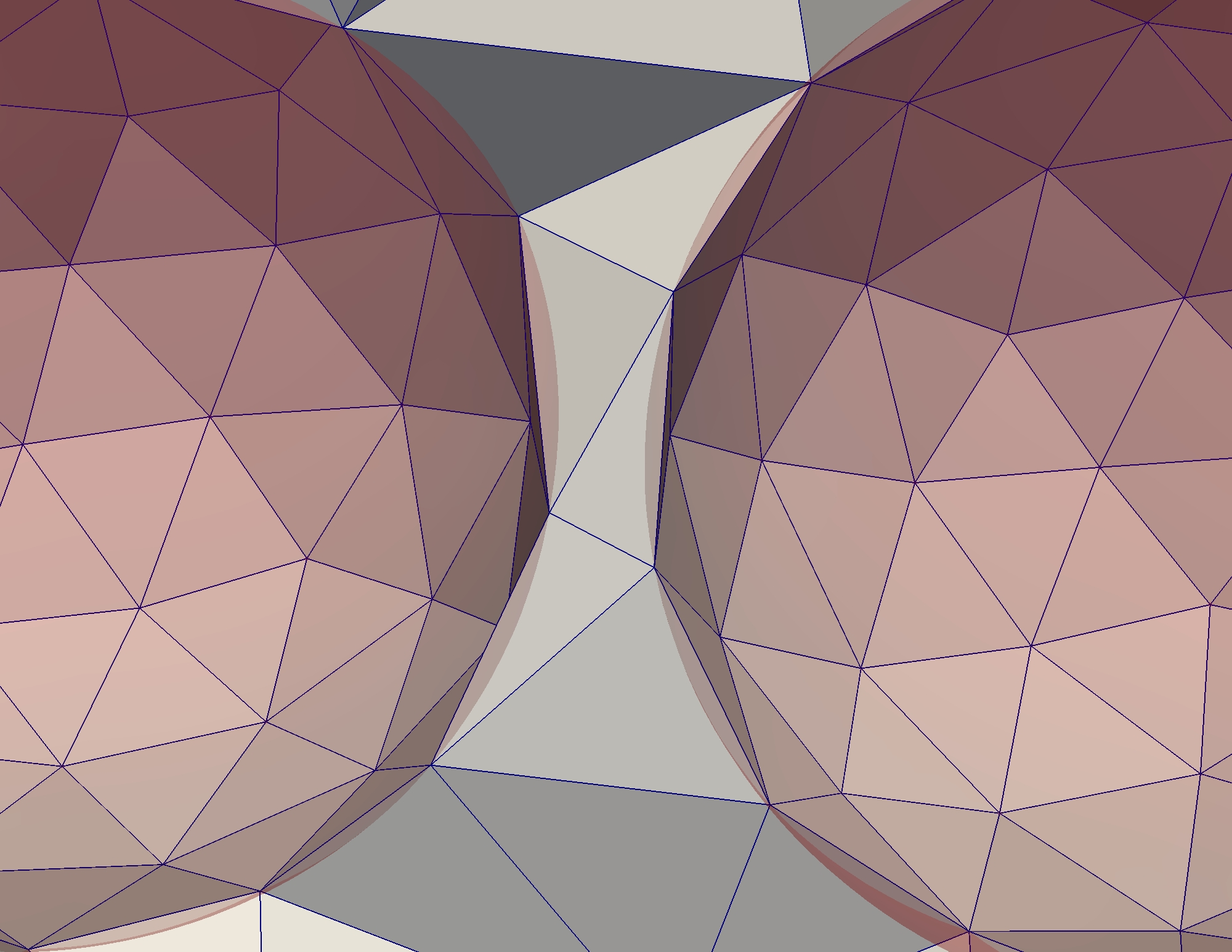}
\hspace*{2em}
\includegraphics[width=0.4\textwidth]{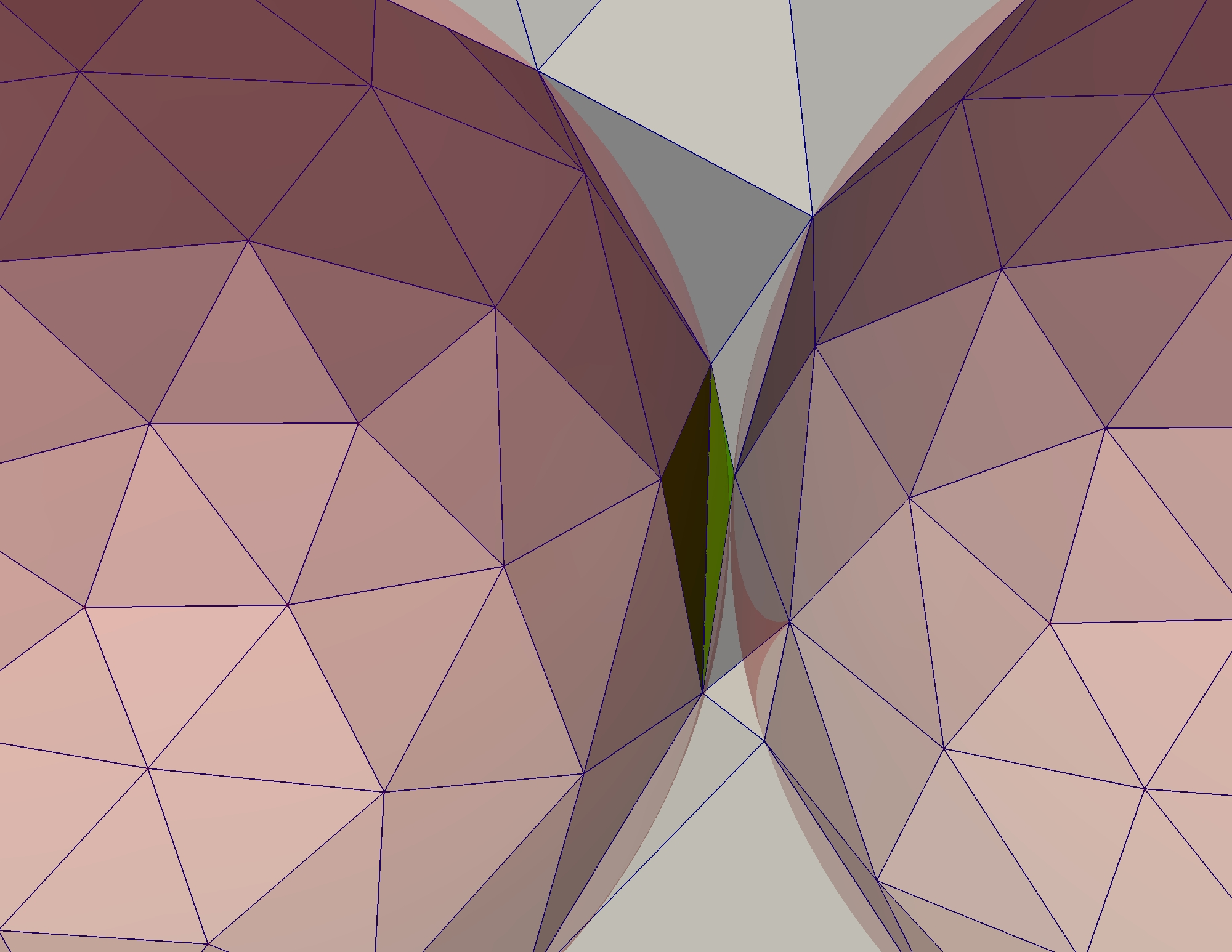}
\caption{Local meshes with different distances $d$ between the spheres. Mesh for the case of a distance of $d = 0.16\ r$ (left) and for $d = 5.31 \cdot 10^{-4}\ r$ (right). In the latter case a degenerated tetrahedron is created (green).}
\label{F:mesh}
\end{figure}

We consider two meshes with $n_s = 214$ spheres and $n_s = 305$ spheres, where the mesh generation is performed using NETGEN \cite{schoeberl_1997}.
For the first mesh ($n_s = 214$), we enforce a minimal distance between the
spheres of $d = 0.16\ r$, while for the second mesh ($n_s = 305$) the
minimal distance is $d = 5.31 \cdot 10^{-4}\ r$, and thus degenerated tetrahedrons
can occur, see Fig.~\ref{F:mesh}.
This can have negative impact on the convergence rate of a standard multigrid method,
but this can be circumvented by appropriate line and plane relaxations in the
block structured grid, see \cite{gmeiner-gradl-gaspar-ruede_2013}.

Any external forces are neglected for this simulation which results in a zero right-hand side $\mathbf{f} = \mathbf{0}$ and the viscosity is assumed to be $\nu = 1$. On the left-hand side of the domain, cf. Fig.~\ref{F:spheres_geom}, a parabolic inflow is prescribed by $\mathbf{g} = ( (1 - x_2^2 - x_3^2)^{1/2}, 0, 0)^\top$, on the right side of the domain a typical outflow (do-nothing condition) is considered, i.e.,
\begin{align*}
T(\mathbf{u}, p)\, \mathbf{n} = \mathbf{0},
\end{align*}
and on the rest of the cylinder surface no-slip boundary conditions, i.e., $\mathbf{u} = \mathbf{0}$, are presumed. On the boundaries of the spheres, we consider either no-slip boundary conditions or free-slip boundary conditions, see, e.g., \cite{pironneau_1989, verfuerth_1987}, where the latter one postulate no flow through the boundary and a homogenous tangential stress, i.e.,
\begin{align*}
\mathbf{u} \cdot \mathbf{n} = 0, \qquad T(\mathbf{u}, p)\, \mathbf{n} \cdot \mathbf{t} = 0,
\end{align*}
where $\mathbf{t}$ denotes the tangential vector.

Let us briefly comment on the implementation of free-slip boundary conditions. While these boundary conditions are straightforward to implement in the case of a plane which is aligned along a coordinate axis, it is more involved in the case of curved boundaries, see, e.g., \cite{engelman-sani-gesho_1982,verfuerth_1987} and more recently \cite{urquiza-garon-farinas_2014}. We realize the homogenous tangential stress by the condition $ (I - \mathbf{n}\, \mathbf{n}^\top )\, T(\mathbf{u}, p)\, \mathbf{n} = \mathbf{0}$ in a point-wise fashion. This approach requires the normal vector in a degree of freedom, where it is particularly important to construct the normals in such a way that they are not interfering with the mass conservation, see, e.g., \cite{engelman-sani-gesho_1982}. Such a mass conservative definition of the normal vector is given for the node $\mathbf{x}_i$ by
\begin{align}\label{normal_massconserv}
\mathbf{n}_i \coloneqq \| \langle \nabla \varphi_i,1\rangle_\Omega \|_2^{-1} \langle \nabla \varphi_i,1\rangle_\Omega,
\end{align}
where $\varphi_i$ denotes the basis function which corresponds to the $i$-th node on the boundary. Note, the integral has to be understood component wise and that this can be realized by a single matrix vector multiplication $B^\top \mathbf{1}$ with additional proper scaling.

\subsection{Simulation results}
The numerical results are performed with the UMG method, since it is the most efficient with respect to time-to-solution and memory consumption. For the computations, we consider the relative accuracy $\epsilon = 10^{-5}$ and a zero initial vector.
%
%
The initial mesh $\mathcal{T}_{-2}$, representing the domain $\Omega$, consists of approximately $1.3 \cdot 10^5$ tetrahedrons. The initial mesh is then refined 5 times, i.e.,
the coarse grid consists of $8.1 \cdot 10^6$ tetrahedrons and the resulting algebraic system on the finest mesh consists of about $2.7 \cdot 10^9$ degrees of freedom.
\begin{figure}[h!]
\centering
\includegraphics[width=0.49\textwidth]{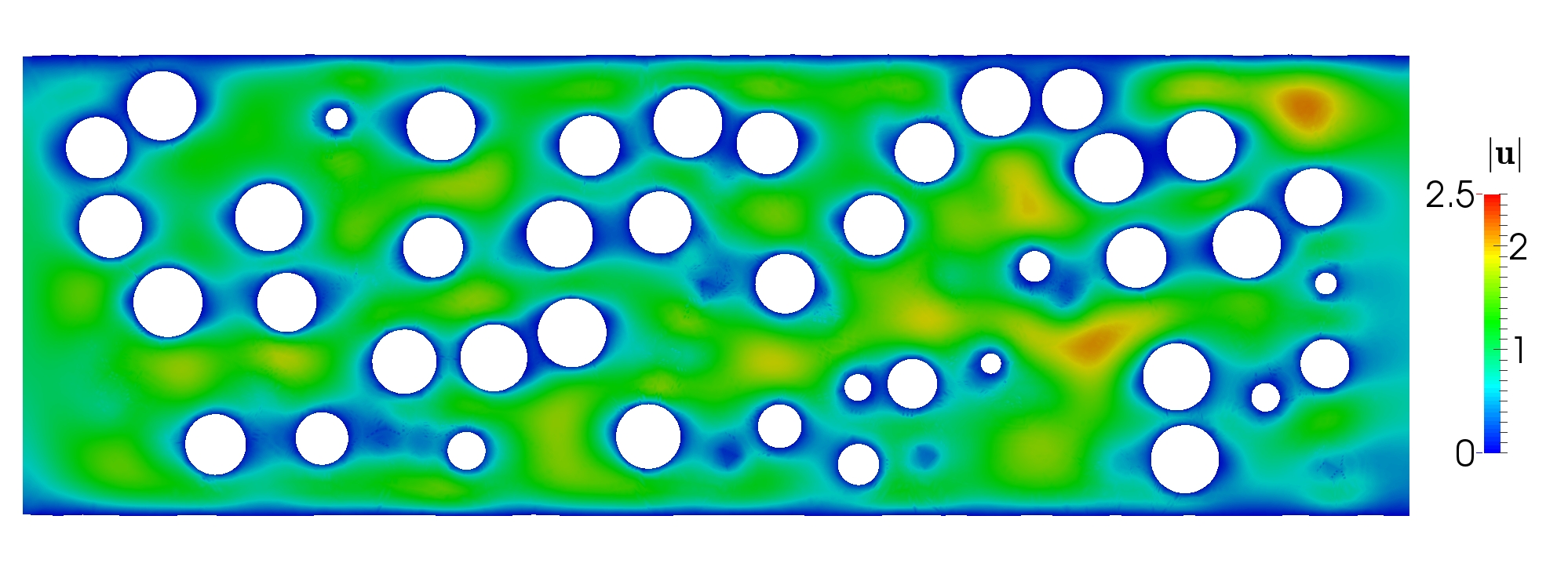}
\includegraphics[width=0.49\textwidth]{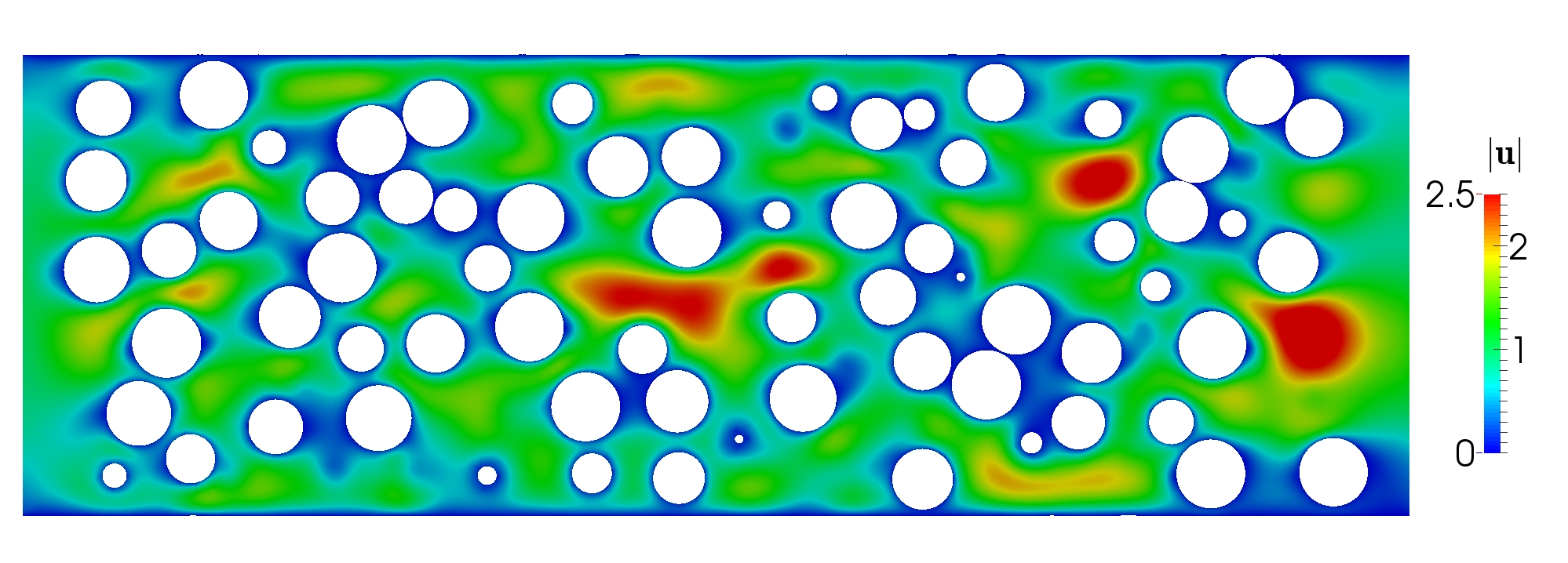}\\
\caption{Slices along the $x_2 = 0$ plane, representing the velocity for $n_s = 214$ spheres (left) and $n_s = 305$ spheres (right).}
\label{F:slices}
\end{figure}
In Fig.~\ref{F:slices}, slices of the velocities along the plane $x_2 = 0$ are presented. We observe higher velocities for the case of $n_s = 305$ spheres, which is due to the denser packing. Note, the computation times are roughly by a factor of 15 faster for the case of $n_s = 214$ spheres (comparable to the times of the previous sections) which can be explained by the more favorable 
shape of the elements of the mesh.
\begin{figure}[!h!]
\centering
\includegraphics[width=0.8\textwidth]{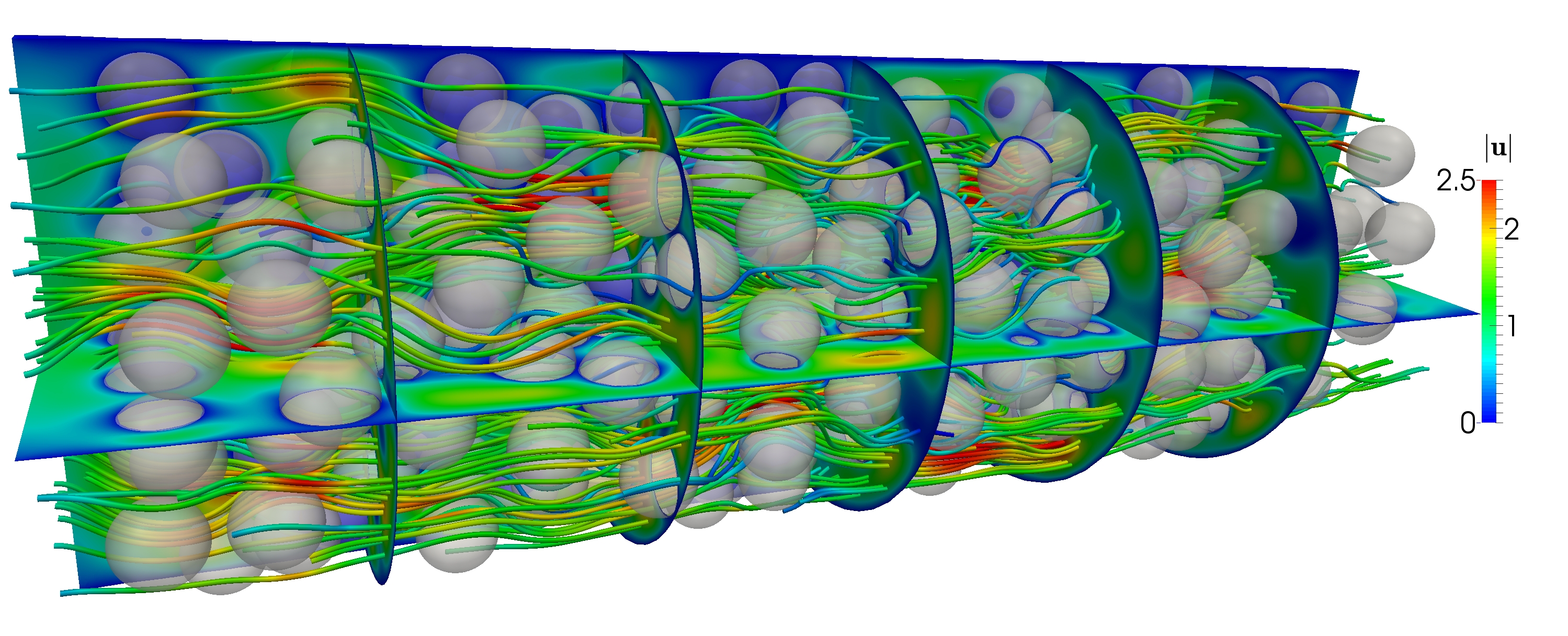}\\
\includegraphics[width=0.8\textwidth]{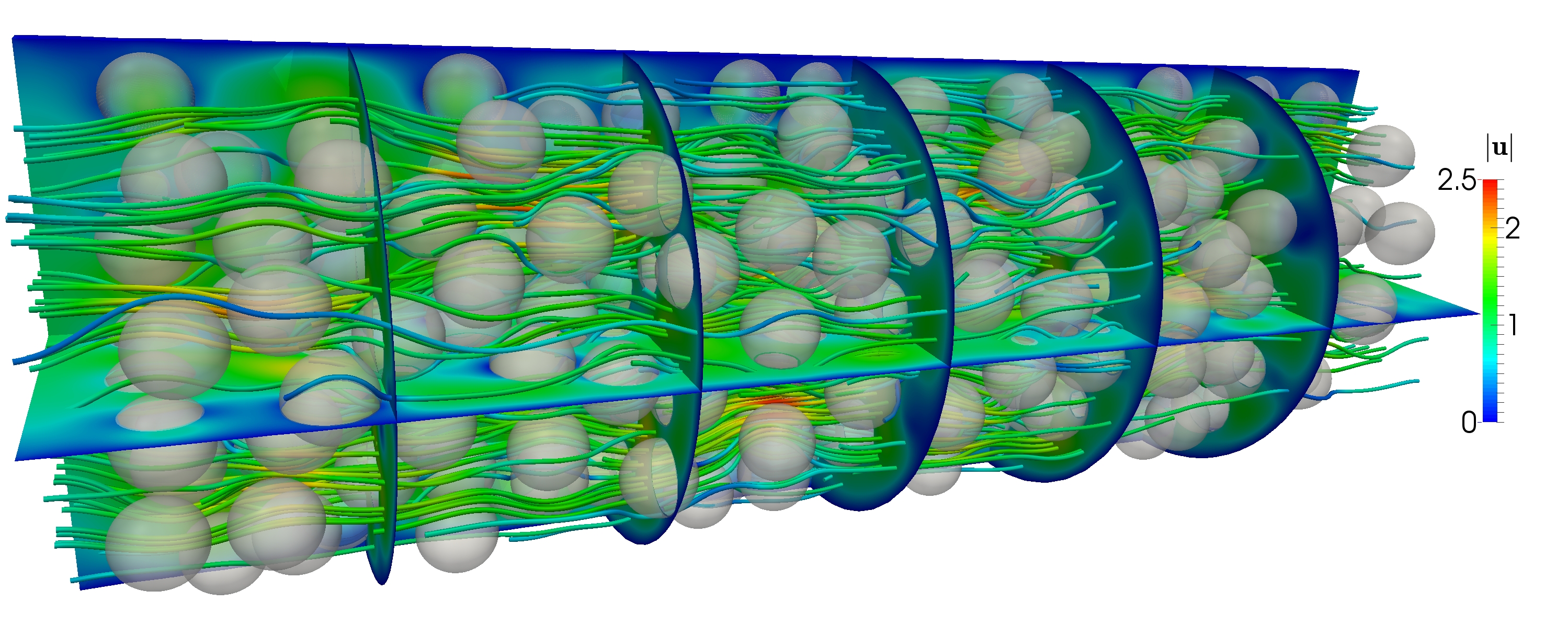}\\
\caption{Streamlines of the velocity, $n_s = 305$ spheres with no-slip bc's (top) and free-slip bc's (bottom) on the spheres surface.}
\label{F:spheres}
\end{figure}
The velocity streamlines for no-slip and free-slip boundary conditions on the spheres are depicted in Fig.~\ref{F:spheres}. We observe that the velocities are higher for the no-slip case, which is due to non-zero velocities at the boundaries of the spheres. The convergence rate of the solver is robust with respect to the different boundary conditions.

\section{Conclusion}\label{sec:conclusion} 
In this article a quantitative performance analysis of three different Stokes solvers is
investigated.
For all methods, robustness and excellent performance for serial execution, and on state-of-the-art parallel peta-scale systems are achieved,
both using the Laplace- and $D$-operator formulation.
These results are explained through a performance analysis, including
the number of operator evaluations. In particular, we investigate the difference 
between the two formulations.
We show that among three state-of-the-art iterative methods, a genuine all-at-once multigrid method (UMG) for the coupled system is the most efficient with respect to time-to-solution and memory consumption. 
This is the key for reaching a system size of
$1.1 \cdot 10^{13}$ DoFs and solving it in less than 13 minutes on 
a current high performance system.
Excellent performance in absolute terms and their scalability makes these methods 
interesting candidates  for reaching the future exascale era. 
Furthermore, we extend the concept of parallel textbook multigrid efficiency 
to the all-at-once multigrid solver
and demonstrate the robustness and efficiency for computing
the flow in a complex domain.

\section*{Acknowledgement}
This work was supported (in part) by the German Research Foundation (DFG) through the Priority Programme 1648 "Software for Exascale Computing" (SPPEXA). The authors gratefully acknowledge the Gauss Centre for Supercomputing (GCS) for providing computing time through the John von Neumann Institute for Computing (NIC) on the GCS share of the supercomputer JUQUEEN at J\"ulich Supercomputing Centre (JSC). The last three authors gratefully acknowledge the hospitality and support of the
Institute for Mathematical Sciences of the
National University of Singapore where part of this work was performed.




\end{document}